\documentclass{elsart}
\usepackage{amsfonts}
\usepackage{amssymb}
\usepackage{graphicx}
\newcommand{\invfpi}{\frac{1}{4\pi}}
\newcommand{\invepi}{\frac{1}{8\pi}}

\begin{document}
\begin{frontmatter}
\title{An Unsplit, Cell-Centered Godunov Method for Ideal MHD}
\author[label1,label3]{Robert K. Crockett\corauthref{cor1}}
\corauth[cor1]{Corresponding author.}
\ead{rcrockett@astron.berkeley.edu}
\ead[url]{astron.berkeley.edu/$\sim$mookie/} \author[label3]{Phillip
Colella} \author[label4]{Robert T. Fisher}
\author[label2,label4]{Richard I. Klein}
\author[label1,label2]{Christopher F. McKee} \address[label1]{Physics
Department, University of California, Berkeley, CA 94720, USA}
\address[label2]{Astronomy Department, University of California,
Berkeley, CA 94720, USA} \address[label3]{Lawrence Berkeley National
Laboratory, Berkeley CA, 94720, USA} \address[label4]{Lawrence
Livermore National Laboratory, Livermore CA, 94551, USA}

Received 27 October 2003; accepted ***

\begin{abstract}
We present a second-order Godunov algorithm for multidimensional,
ideal MHD. Our algorithm is based on the unsplit formulation of
Colella (J. Comput. Phys. vol. 87, 1990), with all of the primary
dependent variables centered at the same location. To properly
represent the divergence-free condition of the magnetic fields, we
apply a discrete projection to the intermediate values of the field at
cell faces, and apply a filter to the primary dependent variables at
the end of each time step. We test the method against a suite of linear
and nonlinear tests to ascertain accuracy and stability of the scheme
under a variety of conditions.  The test suite includes rotated planar
linear waves, MHD shock tube problems, low-beta flux tubes, and a
magnetized rotor problem. For all of these cases, we observe that the
algorithm is second-order accurate for smooth solutions, converges to
the correct weak solution for problems involving shocks, and exhibits
no evidence of instability or loss of accuracy due to the possible
presence of non-solenoidal fields.

\end{abstract}

\end{frontmatter}

\section{Introduction}
In this paper we present a new Godunov method for the equations of
multidimensional ideal magnetohydrodynamics (MHD). We give results
from an implementation of the unsplit, second-order method of Colella
\cite{c90} for these equations. The base scheme solves the ideal MHD
equations using a second-order predictor-corrector formalism. To the
base scheme we add three algorithmic components, whose effects upon
accuracy and stability are measured. The first component is a MAC
projection \cite{bch91,hw65} step, which uses a Poisson solver to
ensure that the cell-edge centered fields used to calculate fluxes are
divergence-free to solver tolerance. This component, though
commonplace in the context of incompressible Navier-Stokes simulation,
is new to the MHD community. The second component is an approximate
projection \cite{abs96} that uses another solution of the Poisson
equation to ensure that the cell-centered field is divergence-free to 
second-order. The last component is a filter \cite{m87} that also
acts to suppress monopole sources in the cell-centered field.
\par
The section that follows covers recent work and some of the schemes
used for ideal MHD simulation. It also introduces methods for
enforcing the divergence-free constraint, and introduces some issues
surrounding multidimensional MHD. Section \ref{algorithm} introduces
our basic algorithm and the extensions we have implemented. A suite of
linear and nonlinear test problems will be used to determine which of
our algorithmic extensions are best suited to each problem type. These
tests and results are covered in Section \ref{tests}. The overall
purpose is to find one combination of these extensions that is well
suited to all of the problems considered. This will be done through
comparisons to published results and in some cases to an eight-wave
MHD algorithm we have implemented. We find that a projection step is
indeed required for accuracy and stability of the schemes. For the
eight-wave scheme in particular, use of the MAC projection is
essential to obtain correct MHD shock jumps. 

\section{Background}
\label{background}
The study of numerical algorithms for magnetohydrodynamics simulations
remains an active one, with no one method having become the
standard. Two generic algorithms are the most widely used at present:
the Method of Characteristics/Constrained Transport (MOC/CT; common in
the astrophysics community) \cite{eh88,sn92} and shock-capturing
(Godunov) methods \cite{bs99,bw88,dw94a,dw94b,p94,rjf95,t00,zmc94}. 
Each has distinct benefits and drawbacks. Codes implementing the
MOC/CT algorithm are relatively simple in design, and satisfy the
divergence-free constraint to machine precision. However, the method
of characteristics used by the ZEUS scheme, as outlined in
\cite{sn92}, is by construction second-order on Alfv\'en and advective
waves, but does not address the two compressive waves of ideal
MHD. Moreover, Falle \cite{f02} found that ZEUS exhibits spurious
rarefaction shocks in certain 1-D MHD shock tubes for a non-isothermal
equation of state. Codes implementing the shock-capturing algorithm on
the other hand, while more complex, give highly accurate results even
for strong shocks. They suffer from the drawback that the
divergence-free constraint is only satisfied to truncation error,
which can be large in the region of large jumps. In order to treat
this difficulty, a variety of techniques have come into use. One such
is the hybrid CT/shock-capturing scheme 
\cite{bs99,dw98,ldz00,rmjf98,t00}, for which the constraint is
satisfied by design like in the MOC/CT case. The cost to the accuracy
of the underlying shock-capturing scheme is unclear. Another approach,
originally due to Brackbill and Barnes \cite{bb80} and implemented by
workers such as Ryu et al \cite{rjf95}, uses a divergence cleaning
step on the cell-centered fields to enforce the constraint. In a third
approach, Powell and co-workers \cite{p94,p99} use a
eight-wave reformulation of the ideal MHD equations originally
due to Godunov \cite{g72}. T\'oth \cite{t00} (hereafter T00)
implements all three types of schemes, among others, using them as the
basis for a comparison on a variety of 1-D and 2-D tests. More
recently, Dedner et al \cite{d02} compare several hyperbolic schemes
with additional waves and divergence-damping terms on the T00 tests.
\par
The 2-D tests of T00 serve to underscore the importance of using
multidimensional problems in evaluating different algorithms, since it
was mainly in this context that differences between them became
apparent. This is to be expected, since errors due to non-solenoidal
fields will generally only show up for problems in two or more
dimensions. Many shock-capturing MHD schemes use an operator-split (or
dimensionally-split) formalism to treat multidimensional ideal
MHD. This means that, for each spatial dimension of the scheme, the
one dimensional MHD equations are applied once. Unsplit schemes, which
instead use the full multidimensional version of the equations, have
been implemented and shown to give results equivalent to those of
split methods for hydrodynamical problems \cite{c90}. Unsplit
shock-capturing schemes for multidimensional ideal MHD are relatively
new, however. One of our main goals is to assess the efficacy of
different approaches for enforcing the divergence-free constraint in
one such unsplit scheme.

\subsection{The Divergence-free Constraint}
Some means must be employed to ensure that the field satisfies the
divergence-free constraint, since this is only guaranteed to within
truncation error in shock-capturing schemes. A dramatic example of
even small errors in the divergence-free condition leading to
instability is given in Section \ref{waves}. There, small amplitude,
non-propagating waves become unstable unless non-solenoidal fields are
smoothed. This phenomenon can be explained in terms of a modified
equation analysis, discussed in Appendix \ref{def}. The possibility of
incorrect field topologies, incorrect dynamics, and numerical
instability motivate efforts to formulate and understand different
means of supressing non-solenoidal errors in the magnetic field.
\par
The eight-wave MHD algorithm, as implemented by Powell et al
\cite{p94,p99} and others, addresses the problem by adding additional
terms corresponding to monopoles to the ideal MHD equations. The
resultant equations are symmetrizable, so that they are Galilean
invariant and transport $\nabla\cdot\vec{B}$ \cite{g72}. The additional
terms show up in two ways for shock-capturing schemes. Since they
modify the 1-D MHD equations used for characteristic tracing to
include an additional eighth wave that travels at the flow speed,
monopoles will be advected along with the flow. Such monopoles could
be carried out of the domain, or they might build up at a stagnation
point. Secondly, the additional terms appear as source terms, making
the system non-conservative if the divergence-free condition isn't
already satisfied. 
\par
Dedner et al \cite{d02} test a scheme that extends the eight-wave
concept to damp and advect monopole sources, even at stagnation points
in the flow. This is done through the magnetic analog of an artificial
compressibility term, an approach that appeared earlier in the context
of the Maxwell equations \cite{m87}. To the extent that the computed
auxiliary field remains continuous, the scheme will remain divergence
free. Any monopole sources will be advected at the fastest speed
allowed under the Courant condition, and damped as they are
advected. This method is very useful on unstructured grids, where
solving the Poisson equation in order to project out the solenoidal
component of the field is difficult.
\par
Divergence cleaning, or Hodge projection, in shock-capturing schemes
can address the problem of non-solenoidal fields in two ways. The most
widely discussed \cite{b98,bb80,rjf95,t00} involves projection of the
cell-centered field onto the space of divergence-free
fields. Projecting in this manner with a centered difference
approximation to the divergence is consistent with the underlying
cell-centered scheme. One is left with fields at the advanced time
which are divergence-free to machine accuracy. Such a projection has
been found to give correct field topologies in shock tube problems
\cite{rjf95}. On the other hand, choosing to eliminate the divergence
as measured in one metric does not guarantee that unphysical effects
are not entering into the dynamics. An illustrative example from
incompressible flow \cite{l93} where an analogous constraint on the
fluid velocity, $\nabla\cdot\vec{v}=0$, occurs shows that checkerboard
modes in the velocity can cause instabilities when using a centered
difference to approximate the divergence. Such modes must be damped by
a suitably chosen filter in order to regain stability.
\par
Another option for cleaning of non-solenoidal fields is to project the
fields at cell-edges, which are then used to calculate fluxes. In a
MAC discretization \cite{bch91,hw65}, the divergence-free constraint
is enforced to within solver tolerance, which can be set as low as
machine precision. We use a multigrid solver to solve the associated 
Poisson equation. The MAC projection has the advantage that it does
not affect the conservation properties of the scheme. 

\subsection{Multidimensional MHD}
The MHD equations in two or more dimensions are decidedly more complex
to solve than in one dimension. In MHD simulations with variations
along the x-axis alone, there is no change in the field along the
x-axis. The divergence-free constraint is therefore trivially
satisfied. Obtaining the solution to the Riemann problem upon which
Godunov methods are based is also relatively straightforward in 1-D.
\par
In multiple dimensions, we are solving the full equations of ideal
MHD, which in conservation form are
\begin{eqnarray}
\label{mhds}
\partial_t\rho + \nabla\cdot(\rho \vec{u}) &=& 0 \\
\partial_t(\rho\vec{u}) + \nabla\cdot\left[ \rho\vec{u}\vec{u} + \left(P +
\frac{B^2}{8\pi}\right)\mathbf{I} - \invfpi\vec{B}\vec{B} \right] &=& 0 \\
\partial_t(\vec{B}) + \nabla\cdot\left[\vec{u}\vec{B} - %
\vec{B}\vec{u}\right] &=& 0 \\
\partial_t(\rho E) + \nabla\cdot\left[ \left(\rho E + P + \invepi %
B^2\right)\vec{u} - \invfpi(\vec{u}\cdot\vec{B})\vec{B} \right] &=& 0,
\label{mhde}
\end{eqnarray}
subject to the constraint $\nabla\cdot\vec{B}=0$. Here $\rho$ is the
mass density, $\rho\vec{u}$ the momentum density, $\vec{B}$ the
magnetic field, and $\rho E=\half\rho|\vec{u}|^2 +
\frac{1}{8\pi}|\vec{B}|^2 + \frac{1}{\gamma -1}P$ the total energy
density. The $\partial_t$ notation denotes derivatives with respect to
time.

\section{Equations and Algorithms}
\label{algorithm}
\begin{figure}[!ht]
\begin{center}
\includegraphics[height=0.3\textheight, width=0.5\textwidth,
scale=5.0]{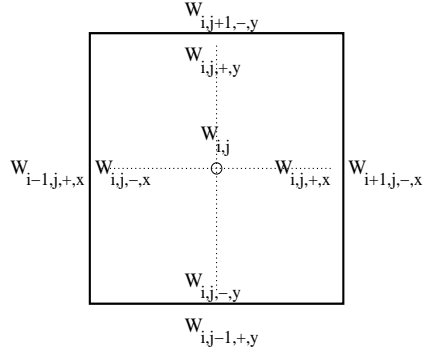}
\caption{Single cell of the computational domain, showing interpolated
states $W^{n+\half}_{i,j,\pm,x}$ and $W^{n+\half}_{i,j,\pm,y}$, and
cell-centered state $W_{i,j}$. The (Riemann) states at cell edges
would lie between the interpolated states. So
$W^{n+\half}_{i+\half,j}$, for instance, would lie between
$W^{n+\half}_{i,j,+,x}$ and $W^{n+\half}_{i+1,j,-,x}$.}
\label{cell}
\end{center}
\end{figure}

In general for hyperbolic conservation laws, the conserved variables
$U$ evolve according to $\partial_tU + \nabla\cdot\vec{F}(U) = 0$. Our
code uses a second-order Godunov-type method for hyperbolic
conservation laws \cite{bct89}. Such volume-average schemes follow the
flux of conserved quantities such as momentum into and out of each
cell comprising the computational domain. Quantities are stored at
cell-centers, their value at this point being the volume-average over
the entire cell. During the course of a timestep, the flux of the
conserved quantities at each edge of each cell is
computed. Differencing these fluxes gives the update of the conserved 
quantity to the next time (see Figure \ref{cell}).
\par
More specifically, our code is based on the unsplit MHD algorithm of
Colella \cite{c90}. It ostensibly tracks the evolution of the
conserved density $\rho$, three components of momentum $\rho\vec{u}$,
three components of magnetic field $\vec{B}$, and total energy density
$\rho E$, in two spatial dimensions. These variables would then be
evolved according to the ideal MHD equations
(\ref{mhds}-\ref{mhde}). However, for simplicity and accuracy a set of
primitive variables $W$ consisting of the density, velocity, field,
and pressure are evolved in time:
\begin{equation}
W=[\rho\; u\; v\; w\; B_x\; B_y\; B_z\; P].
\end{equation}
We will switch freely between $U$ and $W$ here depending on which is
most convenient; the indices will indicate the centering of the
variables. Thus $i,j$ will denote cell-centered states, $i,j,\pm$
forward- and backward-interpolations to cell-edges, and $i+\half,j$
and $i,j+\half$ (Riemann) states at the edges.
\par
The fundamental aspects of our scheme are as follows. Given the state
$W^n_{i,j}$ at time $n$ and spatial coordinate $(x_i,y_j)$, we
simultaneously interpolate in space and extrapolate in time to obtain
the states $W^{n+\half}_{i,j,\pm,x}$ on the two x-boundaries of
each cell $(i,j)$. The same is done for the states at the
y-boundaries, $W^{n+\half}_{i,j,\pm,y}$. All this is done in the normal
and transverse predictor steps of the algorithm. Next, the Riemann
problem is solved using the $\hat{W}^{n+\half}_{i,j,\pm,x}$
states. These Riemann states are used to calculate fluxes at each edge
of a cell in the corrector step. The fluxes are then differenced to
give the update to the next timestep, $W^{n+1}_{i,j}$. We go into more
detail on these steps of our scheme in Sections \ref{npred} through
\ref{riemann}.
\par
The additional parts of our MHD scheme consist of additional terms in
the normal predictor step to ensure correct multidimensional 
behavior, and several steps that address the divergence-free
constraint. These are also discussed in Sections \ref{npred} through
\ref{riemann}.

\subsection{Normal Predictor}
\label{npred}
The predictor computes $W^{n+\half}_{i,j,\pm,x}$ and
$W^{n+\half}_{i,j,\pm,y}$ using a Taylor series expansion:
\begin{equation}
\label{primEvol}
W^{n+\half}_{i,j,\pm,x} = W^n_{i,j} \pm \frac{\Delta x}{2}\partial_xW %
- \frac{\Delta t}{2}\mathbf{A}^x\partial_xW - \frac{\Delta t}{2} %
\frac{\partial U}{\partial W} \partial_yF^y.
\end{equation} 
Here the matrix $\mathbf{A}$ is related to the flux $F^x$ of of the
conserved variable $U$ by $\mathbf{A} = (\partial_U W)(\partial_W
F^x)$. We give formulae for $W^{n+\half}_{i,j,\pm,x}$, those for
$W^{n+\half}_{i,j,\pm,y}$ being similar. The first three terms on the
right-hand side are computed in the normal predictor, and we label
this intermediate result $\hat{W}^{n+\half}_{i,j,\pm,x}$. We now
separate out the evolution of the normal field $B_n=B_x$ through the
following notation:
\begin{equation}
\label{lagrSplit}
\hat{W} = \left[\begin{array}{c}
\tilde{W} \\
B_x \end{array}\right],  \;\;\; 
\mathbf{A} = \left[\begin{array}{cc}
\tilde{\mathbf{A}} & a_B \\
0 & 0 \end{array}\right].
\end{equation}
The matrix $\tilde{\mathbf{A}}$ corresponds to the usual 1-D MHD
equations, with its seven characteristics: forward- and
backward-propagating fast, slow, and Alfv\'en, plus the advective
wave. Second-order accuracy is achieved in part through the use of
characteristic analysis to calculate second-order accurate derivatives
$\partial_xW$ in the spatial interpolation. This characteristic
interpolation is based on calculation of the eigenvalues $\lambda_k$
and left- and right-eigenvectors $l_k$ and $r_k$ of the matrix
$\tilde{\mathbf{A}}$, giving the following expression for the
interpolation of the $\tilde{W}$ variables to cell-edges:
\begin{eqnarray}
\label{charInterp}
\tilde{W}^{n+\half}_{i,j,\pm,x} = \tilde{W}^n_{i,j} + %
\frac{1}{2}\sum_{k: \lambda_k\gtrless 0}\left(\pm 1 - \frac{\Delta %
t}{\Delta x}\lambda_k\right)\alpha_kr_k \\
\alpha_k = \left\{\begin{array}{ll}
{\rm Min}(\alpha^0,\alpha^+,\alpha^-) & {\rm if} \;\alpha^+\alpha^-\ge 0\\
0 & {\rm otherwise}\end{array}\right. \\
\alpha^0=\half l_k\cdot(W^n_{i+1,j}-W^n_{i-1,j}) \\
\alpha^+=2l_k\cdot(W^n_{i+1,j}-W^n_{i,j}) \\
\alpha^-=2l_k\cdot(W^n_{i,j}-W^n_{i-1,j})
\end{eqnarray}
The $\alpha_k$ represent the strength of the k$^{\rm{th}}$ wave in the
interpolant. The sum over $\lambda_k<0$ would correspond to
backward-propagating waves used in the interpolation to
$\tilde{W}^{n+\half}_{i,j,-,x}$, and similarly for $\lambda_k>0$ and
$\tilde{W}^{n+\half}_{i,j,+,x}$.
\par
A full accounting for all x-derivative terms in the 2-D MHD equations
shows that $a_B$ is given by
\begin{equation}
\label{stoneTerms}
a_B = -\left[0,\;\frac{B_x}{4\pi\rho},\;\frac{B_y}{4\pi\rho},\; %
\frac{B_z}{4\pi\rho},\; v,\; w,\; %
\frac{\vec{u}\cdot\vec{B}}{4\pi}\right]^T.
\end{equation}
These terms are essential to the second-order accuracy of the scheme,
in particular on multidimensional problems such as waves not
propagating along the coordinate axes. They are incorporated into our
algorithm through a simple finite differencing of the normal
derivative of the normal field. The terms are added to those already
present due to the characteristics-based interpolation: 
\begin{equation}
\tilde{W}^{n+\half}_{i,j,\pm ,x} := \tilde{W}^{n+\half}_{i,j,\pm,x} - %
\frac{\Delta t}{2}a_B(\mathcal{D}^0_xB_x^n)_{i,j}.
\end{equation} 
The $(\mathcal{D}^0_xB_x^n)_{i,j} =
((B_x)^n_{i+1,j}-(B_x)^n_{i-1,j})/(2\Delta x)$ correction term is the
centered-difference approximation to $\partial_xB_x$. Note that this
approximation to the derivative is not limited. The need for such
correction terms in an unsplit scheme for multidimensional MHD was not
addressed in C90. It was first noted by Stone \cite{s04} during an
examination of the accuracy of an unsplit Godunov scheme on advected
flux rings. 

\subsection{Transverse Predictor}
\label{tpred}
The last term in the evolution equation (\ref{primEvol}) is included
via the transverse predictor. The basic idea is to approximate
transverse derivatives (in this case in the y-direction) using a 1-D
Godunov method. We take the states calculated in the normal predictor,
$\hat{W}^{n+\half}_{i,j,\pm ,y}$, and first use them to solve the
Riemann problem at each y-boundary in the domain. The resulting
Riemann states $U^{n+\half}_{i,j+\half}$ are subsequently used to
calculate the fluxes needed for the last term in equation
\ref{primEvol}. In more formal terms,
\begin{eqnarray}
U^{n+\half}_{i,j+\half} = %
\mathcal{R}\left(\hat{W}^{n+\half}_{i,j,+,y}\, ,\; %
\hat{W}^{n+\half}_{i,j+1,-,y}\right) \\
F^{y,n+\half}_{i,j+\half} = F^y\left(U^{n+\half}_{i,j+\half}\right).
\end{eqnarray}
Here $\mathcal{R}(\cdot,\cdot)$ denotes the Riemann problem solution
using the two states on either side of an edge as input; see Section
\ref{riemann}. It is then a straightforward matter to use these fluxes
to calculate a finite-difference approximation to $\partial_yF^y$, and
thereby complete the calculation of the edge-centered states.

\subsection{Corrector}
\label{corr}
The corrector first calculates fluxes at all cell-edges using another
Riemann problem solve. At this stage, we have the first-order accurate
approximation to the interpolated states $W^{n+\half}_{i,j,\pm,\cdot}$
from the predictor in hand. The Riemann solver takes these states and
returns a single state for each cell-edge, $U^{n+\half}_{i+\half,j}$
and $U^{n+\half}_{i,j+\half}$. For instance,
\begin{equation}
U^{n+\half}_{i+\half,j} = \mathcal{R}\left(W^{n+\half}_{i,j,+,x}\, %
,\; W^{n+\half}_{i+1,j,-,x}\right).
\end{equation}
The formula for $U^{n+\half}_{i,j+\half}$ is similar. See Section
\ref{riemann} for more details on the Riemann problem solution.
\par
The states at cell-edges are not guaranteed to be divergence-free. We
modify the C90 algorithm to enforce the divergence-free condition for
the Riemann problem states. These Riemann states have a non-solenoidal
component that we treat through a MAC projection, earlier used in the
context of incompressible fluid computations \cite{bch91}. The
edge-centered fields $(\vec{B}^*)^{n+\half}_{i+\half,j}$ and
$(\vec{B}^*)^{n+\half}_{i,j+\half}$ are used to calculate a
cell-centered monopole charge density $q_M=\nabla\cdot\vec{B}^*$, and
a Poisson solver is in turn used to find the scalar field $\phi$
implied by this monopole charge density distribution. The scalar
field satisfies the following relations, in which
$\mathcal{D}^\pm_{x}$ correspond to the forward- and
backward-difference approximations to the derivative
$\frac{\partial}{\partial x}$, and similarly for
$\mathcal{D}^\pm_{y}$:
\begin{eqnarray}
(q_M)_{i,j} &=& \mathcal{D}^-_x(B^*_x)_{i+\half,j} +
\mathcal{D}^-_y(B^*_y)_{i,j+\half} \\
\label{FDPoisson}
\left[\mathcal{D}^+_x\mathcal{D}^-_x + %
\mathcal{D}^+_y\mathcal{D}^-_y\right]\phi_{i,j} &=& (q_M)_{i,j}.\\
\end{eqnarray}
The correction to the field is calculated from $\phi$ as follows:
\begin{eqnarray}
(B_x)_{i+\half,j} &=& (B^*_x)_{i+\half,j} - \mathcal{D}^+_x\phi_{i,j}
\\ (B_y)_{i+\half,j} &=& (B^*_y)_{i+\half,j} -
\half\left[\mathcal{D}^0_y\phi_{i+1,j} +
\mathcal{D}^0_y\phi_{i,j}\right] \\ (B_x)_{i,j+\half} &=&
(B^*_x)_{i,j+\half} - \half\left[\mathcal{D}^0_x\phi_{i,j+1} +
\mathcal{D}^0_x\phi_{i,j}\right] \\ (B_y)_{i,j+\half} &=&
(B^*_y)_{i,j+\half} - \mathcal{D}^+_y\phi_{i,j}
\end{eqnarray}
With this correction to the magnetic field of the Riemann states, the
$L^1$ norm of the MAC monopole density is reduced from its initial
value by a user-settable multiplicative factor, in our case
$10^{-12}$.
\par
The algorithm now proceeds to calculate the fluxes associated with the
Riemann states. These fluxes are then differenced to give the update
to the next time $U^{n+1}_{i,j}$:
\begin{eqnarray}
F^{x,n+\half}_{i+\half,j} = F^x\left(U^{n+\half}_{i+\half,j}\right) \\
U^{n+1}_{i,j} = U^n_{i,j} - \Delta
t\,\mathcal{D}^-_xF^{x,n+\half}_{i+\half,j} - \Delta %
t\,\mathcal{D}^-_yF^{y,n+\half}_{i,j+\half}
\end{eqnarray}
After the update to $t^{n+1}$, we are left with a cell-centered field
$\vec{B}^{*,n+1}_{i,j}$ that is no longer divergence-free by a
centered-difference divergence metric. To what extent, if any, this is
a problem depends on the physical problem being considered, and will
be addressed later.
\par
For those cases where a reduction of the divergence is required, two
algorithmic extensions have been implemented. The first follows from
noting that it is desirable to have a diffusive term of the form (see
\cite{m87,d02}):
\begin{equation}
\frac{\partial(\nabla\cdot\vec{B})}{\partial t} = %
\eta\nabla^2(\nabla\cdot\vec{B}) 
\end{equation}
act on the divergence of $\vec{B}$. This may be rewritten to eliminate
a spatial derivative by pulling out a divergence operator, giving
\begin{equation}
\label{divDiff}
\frac{\partial\vec{B}}{\partial t} = \eta\nabla(\nabla\cdot\vec{B})
\end{equation}
A simple, single-step filter may be derived \cite{mc01} as a
finite-difference approximation of equation \ref{divDiff}:
\begin{eqnarray}
\label{filterx}
B_x := B_x + \eta\Delta t\,(\mathcal{D}^+_x\mathcal{D}^-_xB_x + %
\mathcal{D}^0_x\mathcal{D}^0_yB_y) \\
\label{filtery}
B_y := B_y + \eta\Delta t\,(\mathcal{D}^0_x\mathcal{D}^0_yB_x + %
\mathcal{D}^+_y\mathcal{D}^-_yB_y)
\end{eqnarray}
The advantage of this choice of discretization of Equation
\ref{divDiff} lies in its effective damping of checkerboard modes; see
\cite{r98} for analysis and comparison of different filtering
schemes. 
\par
In order to choose a value for $\eta$ we use Fourier stability
analysis, giving a stability condition for the scheme in equations
\ref{filterx} and \ref{filtery} of $\Delta t \leq \frac{2(\Delta
x)^2}{5\eta}$. Since $\Delta t$ is set by the Courant condition, we
are able to derive a condition on $\eta$, giving the maximum amount of
diffusion of the monopole sources possible, given our timestep and
grid spacing:
\begin{equation}
\eta = C\frac{(\Delta x)^2}{\Delta t}
\end{equation}
with $C\leq\frac{2}{5}$. A stronger condition, $C\leq\frac{1}{5}$,
will always damp monopole modes. This formulation will both decrease
the cell-centered divergence and damp checkerboard modes. The
filtering is always used when an approximate projection is performed.
\par
The numerical effect of the filter can be modified through the
parameter $C$. We found, through the nonlinear tests outlined in
Section \ref{tests}, that values in the range $10^{-2}-10^{-1}$ worked
best. We note here that we chose to apply the filter to the conserved
variables $U^{n+1}$, so that the changed magnetic field causes no
change in the total energy. Any addition (subtraction) of magnetic
energy shows up as a decrease (increase) in the internal energy. 
\par
The second means for treating non-solenoidal cell-centered fields is
an approximate projection, in which the cell-centered divergence is
used to calculate a monopole charge density $q_M =
\mathcal{D}^0_xB^*_x + \mathcal{D}^0_yB^*_y$. Note that the multigrid
Poisson solver uses the finite-difference operator
$\mathcal{D}^+\mathcal{D}^-$, not $\mathcal{D}^0\mathcal{D}^0$, for
the Laplacian $\nabla^2$. The resulting solution is not discretely
divergence-free, as it would be if we used the operator
$\mathcal{D}^0\mathcal{D}^0$ \cite{abs96}, but instead second-order
accurate. However, the extension of exact solvers to adaptive meshes
is extremely complicated \cite{hb97}, whereas the approximate
projection used here is straightforward to extend to AMR
\cite{mc00}. We note moreover that use of a centered-difference
measure of divergence leaves unchanged unphysical checkerboard
modes \cite{l93,mc01}. An extensive analysis of approximate
projections is given in \cite{abc00}.
\par
The Poisson solve yields the scalar field $\phi$ satisfying equation
\ref{FDPoisson}. We difference $\phi$ as follows to give the corrected
field:
\begin{eqnarray}
B^{n+1}_x = B^{*,n+1}_x - \mathcal{D}^0_x\phi_{i,j} \\
B^{n+1}_y = B^{*,n+1}_y - \mathcal{D}^0_y\phi_{i,j}
\end{eqnarray}
We find in our tests that an approximate projection does not have a
significant impact on the quality of our results when used in
conjunction with a MAC projection. Given the significant computational
expense of using both, we advise against such a scheme.

\subsection{Riemann Solver}
\label{riemann} 
We solve the Riemann problem for ideal MHD using a linearized solver
\cite{t99}. It employs characteristic analysis, like that of the
normal predictor of Section \ref{npred}, to solve for the states at
cell edges. In this case, eigenvalues $\lambda_k$ and eigenvectors
$l_k$ and $r_k$ are calculated using the arithmetic average of the
states to the left and right of the edge. (Call this the base state.)
For example, the solution to the Riemann problem at $x_{i+\half}$,
$W^{RP}_{i+\half}$, is built from the appropriately directed jumps
across waves as follows:
\begin{equation}
W^{RP} = \left\{\begin{array}{rl}
\half\left( W^R - \sum_{k, \lambda_k>0}\alpha_k r_k +%
W^L + \sum_{k, \lambda_k<0}\alpha_k r_k \right), & \textrm{ if }
|u|\leq\epsilon_v \\
W^L + \sum_{k, \lambda_k<0}\alpha_k r_k, & \textrm{ if } u<-\epsilon_v \\
W^R - \sum_{k, \lambda_k>0}\alpha_k r_k, & \textrm{ if } u>\epsilon_v
\end{array}\right.
\end{equation}
Here $|u|$ is the advective velocity (along $\hat{e}_x$ in this case)
of the base state. The constant $\epsilon_v$ is chosen to be
$10^{-14}$ times the largest characteristic speed $\lambda_k$. Note
the averaged solution used for small advective speeds. This ensures
that advective velocities on the order of machine precision, whose
sign is random, do not cause asymmetries in the Riemann problem
solution. Finally, note that in order to maintain consistency, the
longitudinal component of the magnetic field in the solution (in this
case $B^{RP}_x$) is assigned the value of the longitudinal magnetic
field of the base state, $\half (B^L_x+B^R_x)$.
\par
For strongly nonlinear problems, it sometimes happens that the CFL
condition is not sufficient to keep a scheme stable. Ordinarily, we
use a simple minimum of timesteps implied by the one-dimensional CFL
condition. If accelerations are large, velocities can grow such that
in one timestep the pressure or density becomes negative. In order to
dynamically adjust to such situations, we implement a scheme that
checks for negative pressures or densities in the cell-centered
states, $W^{n+1}_{i,j}$, after the corrector step. If one is
encountered, that timestep is restarted with the CFL number lowered by
a factor of two from its nominal value, down to a minimum of
one-eighth of the initial CFL number. Once the code has proceeded for
several timesteps without again encountering negative values, the CFL
number is raised by a factor of two, eventually reaching its initial
value. In practice, negative values are encountered only rarely, so
that the average CFL number over the course of a run is close to the
nominal value. Note that even if the CFL ramps quickly back up to the
nominal value, the timestep is never allowed to increase by more than
10\% in one iteration. In this way, the effect of an increase in CFL
number is in fact spread over several iterations. Note that the
average CFL number remains above 0.4 in all tests of Section
\ref{tests}.

\section{Numerical Tests}
\label{tests}
In this section, we compare the behavior of the code on a variety of
linear and nonlinear problems. Both seven- and eight-wave MHD
codes were used. The code implementing the eight-wave MHD
algorithm also uses a predictor-corrector formalism. The
characteristic analysis performed in the predictor steps uses an
eighth wave carrying changes in the normal field at the advection
velocity, in addition to the seven waves of ideal MHD. As a result, in
equation \ref{lagrSplit} of Section \ref{npred}, the $(2,2)$ entry of
$\mathbf{A}$ is equal to the advective speed $u$ and not zero. The
terms needed for second-order accuracy in ideal MHD (equation
\ref{stoneTerms}) are accounted for in this case, so that $a_B=0$. The
eight-wave algorithm also adds a source term,
\begin{equation}
\label{source}
S=-\left[0,\; \frac{B_x}{4\pi},\; \frac{B_y}{4\pi},\; %
\frac{B_z}{4\pi},\; u,\; v,\; w,\; %
\frac{\vec{u}\cdot\vec{B}}{4\pi}\right]^T(\nabla\cdot\vec{B}),
\end{equation}
to the right-hand side of the MHD equations \ref{mhds}-\ref{mhde}. The
source term is calculated in the transverse predictor and corrector
steps. It is added to the updated states along with the differenced
fluxes. We note, however, that performing a MAC projection guarantees 
that $\nabla\cdot\vec{B}=0$ for the edge-centered states, and as a
result the source term calculated in the corrector is numerically zero
in this case. This implies that only with a MAC projection is the
scheme conservative and satisfies the jump relations. The source term
in the transverse predictor, while generally small, is not numerically
zero.
\par
For both MHD implementations, different variations on the base
algorithm were tested. In what follows, codes with conservative
filtering are labeled with 'CF'. An 'AP' denotes codes with an
approximate projection; it is always accompanied by conservative
filtering, in order to suppress checkerboard modes. Those codes
utilizing a MAC projection are labeled with 'MAC'. So, for example, a
MAC projection code with filter is labeled 'MAC+CF'. 

\subsection{Simple Linearized MHD Waves}
\label{waves}
We tested the performance of the code on all four varieties
(advective, fast, slow, and Alfv\'en) of linearized MHD waves. These
tests comprised waves propagating along x- and y-axes (wavenumbers
$\vec{n}=(1,0)$ and $\vec{n}=(0,1)$), along with waves at slopes of
1:1 ($\vec{n}=(1,1)$) and 2:1 ($\vec{n}=(2,1)$). The simulation domain
had length $L=1$ in both dimensions, and the boundaries in all cases
were periodic. The Alfv\'en waves used $\rho_0=1$, $\vec{u}_0=0$,
$\vec{B}_0=B_0\hat{b}=\sqrt{4\pi}\hat{b}$ with unit vector
$\hat{b}=(\frac{1}{\sqrt{2}},\frac{1}{\sqrt{2}})$, and $P_0=1$. The
perturbation $\delta W$ is
\begin{equation}
\delta W = \left[\begin{array}{c}
0 \\
0 \\
0 \\
-c_A \\
0 \\
0 \\
B_0 \\
0\end{array}\right]\delta_{\rm pert}\sin\left(\vec{k}\cdot\vec{x}\right), \;\;\textrm{with}\;\; \vec{k} = 2\pi\vec{n},
\end{equation}
where $\vec{n}$ is the aforementioned vector of integers chosen so as
to be consistent with the periodic boundaries. The Alfv\'en speed
$c_A=B_0/\sqrt{4\pi\rho_0}=1$.
\par
Fast and slow wave expressions are somewhat more complicated. In this
case, $\hat{b}$ lies at 45 degrees from the unit wavevector $\hat{k}$,
all other aspects of the unperturbed state remaining unchanged from
the Alfv\'en case. The perturbation is
\begin{eqnarray}
\delta W = \left[\begin{array}{c}
\rho_0 \\
\frac{\sqrt{2}c_{F/S}^2\hat{b}_y-a^2\hat{n}_y}{c_{F/S}} \\
\frac{a^2\hat{n}_x-\sqrt{2}c_{F/S}^2\hat{b}_x}{c_{F/S}} \\
0 \\
-\sqrt{2}B_0\frac{c_{F/S}^2-a^2}{c_A^2}\hat{n}_y \\
\sqrt{2}B_0\frac{c_{F/S}^2-a^2}{c_A^2}\hat{n}_x \\
\rho_0a^2\end{array}\right]\delta_{\rm %
pert}\sin\left(\vec{k}\cdot\vec{x}\right) \\
a^2 = \frac{\gamma P_0}{\rho_0} \\
c_{F/S}^2 = \frac{1}{2}\left(a^2 + c_A^2 \pm \sqrt{c_A^4 + a^4}\right) %
\;\;\textrm{corresponding to fast/slow speeds} \\
\hat{n} = (\hat{n}_x,\,\hat{n}_y) = \frac{\vec{n}}{\mid\vec{n}\mid} 
\end{eqnarray}
All tests gave results consistent with second order accuracy, although
waves at 2:1 showed slightly smaller ($\geq 1.8$, versus 2.0) rates of
convergence in some components. The modifications suggested by Stone
were essential in obtaining second-order convergence for the 1:1 and
2:1 slope tests, increasing the convergence rates from first- to
second-order for waves propagating along the diagonal. An example of
the convergence rates for small amplitude fast waves is given in Table
\ref{convTable}.
\par
\begin{table}[!ht]
\begin{center}
\begin{tabular}{|l|c|c|} \hline
Component & Without correction & With correction \\ \hline\hline
$\rho$    & 0.977              & 2.03 \\ \hline
$v_x$     & 1.40               & 2.03 \\ \hline
$v_y$     & 1.15               & 2.02 \\ \hline
$B_x$     & 1.60               & 2.01 \\ \hline
$B_y$     & 1.57               & 2.02 \\ \hline
$P$       & 0.977              & 2.03 \\ \hline\hline
\end{tabular}
\caption{Convergence rates by component for fast waves ($\delta_{\rm
pert}=10^{-5}$) propagating at 45 degrees, for seven-wave MHD code
without and with corrections suggested by Stone \cite{s04}.}
\label{convTable}
\end{center}
\end{table}
A more stringent test of the code is to advect the linearized waves so
that their profile remains stationary. In these tests, the background
state has a non-zero velocity equal to the wavespeed:
\begin{eqnarray}
\textrm{Alfv\'en:}\;\; \vec{u}_0 = -c_A\hat{n} \\
\textrm{Fast/Slow:}\;\; \vec{u}_0 = -c_{F/S}\hat{n}
\end{eqnarray}
An analysis of the eigenstructure of MHD shows (Appendix \ref{def})
that such waves should cause trouble for the seven-wave MHD codes that
do not suppress non-solenoidal fields. Errors generated in this case are
not advected away, causing difficulty in the case that they are not
diffused or otherwise dealt with. We find that the seven-wave MHD
algorithm without the application of any projection or filter is
unstable for low amplitude ($\delta_{\rm pert} = 10^{-4}$) fast waves
with $\hat{n}$ at 2:1 slope. The instability starts as a high
frequency oscillation in the field parallel to $\vec{n}$. It grows 
with time and spreads to the other components, causing a low order of
convergence even at early times. Once the oscillations reach a certain
level, the solution becomes unstable. The oscillations are almost
completely absent, and the code stable, using a MAC projection. Adding
filtering of the field further reduces the errors. A code including
filtering but no MAC projection, and one with approximate projection
and filtering both stabilize the scheme. We conclude that either
projection or filtering is essential to stability for this class of
problems.

\subsection{Decay of Linearized MHD Waves}
The equations of ideal MHD neglect the effects of viscosity and
electrical resistivity. Numerical dissipation, however, can affect the
solution in ways that mirror these physical effects. Following Ryu,
Jones, and Frank \cite{rjf95} (hereafter RJF95), we measure the decay
of Alfv\'en, fast, and slow waves, and use the implied physical
resistivity as a measure of the numerical resistivity of our ideal MHD
scheme.
\par
\begin{figure}[!ht]
\begin{center}
\includegraphics[height=0.5\textheight, width=0.9\textwidth,%
scale=1.2]{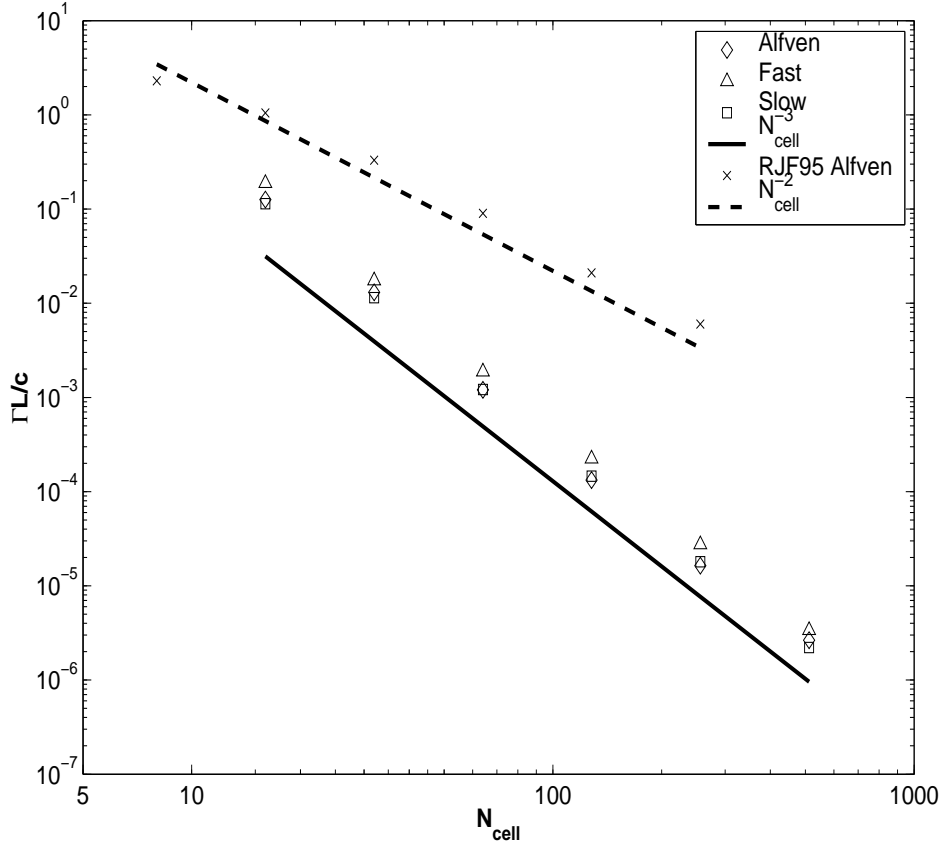}
\caption{Normalized damping rate versus resolution for the decay of
Alfv\'en, fast, and slow modes propagating at a 1:1 slope. The
calculations were done at resolutions of $16\times 16$, $32\times 32$,
etc. up to $512\times 512$.}
\label{wDecay}
\end{center}
\end{figure}

We use exactly the same set-up as RJF95, with Alfv\'en, fast, and slow
waves propagating at a 1:1 slope and wavelength of $\sqrt{2}$ times
the length of one side of the computational domain. The decay of these
waves was measured by fitting a decaying exponential to a measure of
the wave strength,
\begin{equation}
\varepsilon = \sum_{i,j}l_k\cdot(\delta W)_{i,j},
\end{equation}
where $(\delta W)_{i,j}=W_{i,j}-W_0$ is the perturbation in the
primitive variables. $l_k$ is the  left-eigenvector associated with
the mode in question, evaluated at $W_0$, the unperturbed state.
Figure \ref{wDecay} shows the dependence of the decay rate on
resolution for runs with number of cells per dimension $N=$16, 32, 64,
128, 256, and 512. Comparison with Figure 4 of RJF95 reveals a much
smaller decay rate for our unsplit method. Furthermore, we find that
the decay rate varies according to the power law $\Gamma\propto
N^{-3}$, not $N^{-2}$. The former is consistent with second-order
accuracy, since the truncation errors have the following form:
\begin{equation}
\tau =  C_2(\Delta x)^2\frac{\partial^3 U}{\partial x^3} + %
C_3(\Delta x)^3\frac{\partial^4 U}{\partial x^4} + %
\mathcal{O}(\Delta x^4).
\end{equation}
We see here that the second-order error term is proportional to
$\frac{\partial^3 U}{\partial x^3}$, which is dispersive in
nature. The third-order term is the first error term that is
dissipative. As a result, we expect the dissipation to decrease as
$\Delta x^3$, or $N^{-3}$. We have no good explanation for the
$\mathcal{O}(\Delta x^2)$ decay law observed in RJF95, aside from the
possible effects of artificial viscosity. 

\subsection{MHD Shock Tube}
%
%
The second test is the MHD shock tube problem from Ryu, Jones, and
Frank \cite{rjf95}. The solution consists of two fast shocks,
one slow shock, one slow rarefaction, and a contact discontinuity. We
have run the problem in two orientations: (1) with the shock velocity
aligned with the x-axis of the computational domain (referred to below
as 1-D), and (2) with this velocity inclined at a 2:1 slope. The
latter configuration follows the 2-D shock tube test case from T00. It
was chosen in order to test the multidimensional behavior of the code,
meanwhile ensuring there were no serendipitous cancellations of
errors, as might be the case in a $45^\circ$ inclined case. 
In runs of a 1-D tube at $R_{512}$ (ie. 512 cells per linear
dimension) on a domain of size $L=1$ to a time of $t=0.08$, we are
able to reproduce the results given in Table VI of Dai \& Woodward
\cite{dw94a} (hereafter DW94) with errors of $\leq 0.12\%$. These
results are independent of the type of filter used and whether an
approximate projection was performed. Both the seven- and eight-wave
codes give exactly the same result. When combined with an observed
first-order convergence rate, they give us great confidence in the
performance of both codes on 1-D shock tube problems.
\par
The inclined version of this shock tube problem was run on grids of
size $2N\times N$, with $N=64$, 128, and 256. Unlike in the aligned
shock tube, in this case divergence cleaning is required. Since the
natural boundary conditions for the problem would be tedious to
implement in our multigrid solver, we chose instead to embed the
physical domain within a larger one, four times its size. The data in
the region outside of the original one is filled using the natural
symmetry of the problem. The Poisson equation is then solved on this
enlarged domain, using homogeneous Neumann boundary conditions. We have
verified that the field in the original domain is clean to within the
tolerances used previously.
\par
We compared the results from the inclined shock tube with coarsened
versions of a $R_{4096}$ 1-D shock tube results by first taking cuts
of the data that included all cells lying along a line at 2:1 slope
($\alpha = 26.57^\circ$). Then, all velocities and fields in the cut
were rotated by $-\alpha$. A check using the initial conditions showed
that such a cut of the inclined initial conditions matched perfectly
with the 1-D shock tube ICs outside of the jump region. (In the jump
region, the inclined ICs were slightly different due to the volume
averaging performed in producing them. This leads the initial jump to
be spread over two zones instead of one.) Note that we chose the grid
spacing for the inclined runs to be a factor of $\sqrt{5}$ smaller
than the 1-D grid spacing, so that the shock covers the same distance
in physical space in a given time. The domain was therefore of length
$L=1$ in the direction of shock propagation, and the final time
$t=0.08$, in both the aligned and inclined shock runs. 
\begin{figure}[!ph]
\begin{center}
\includegraphics[height=0.92\textheight, width=0.92\textwidth,%
scale=1.0]{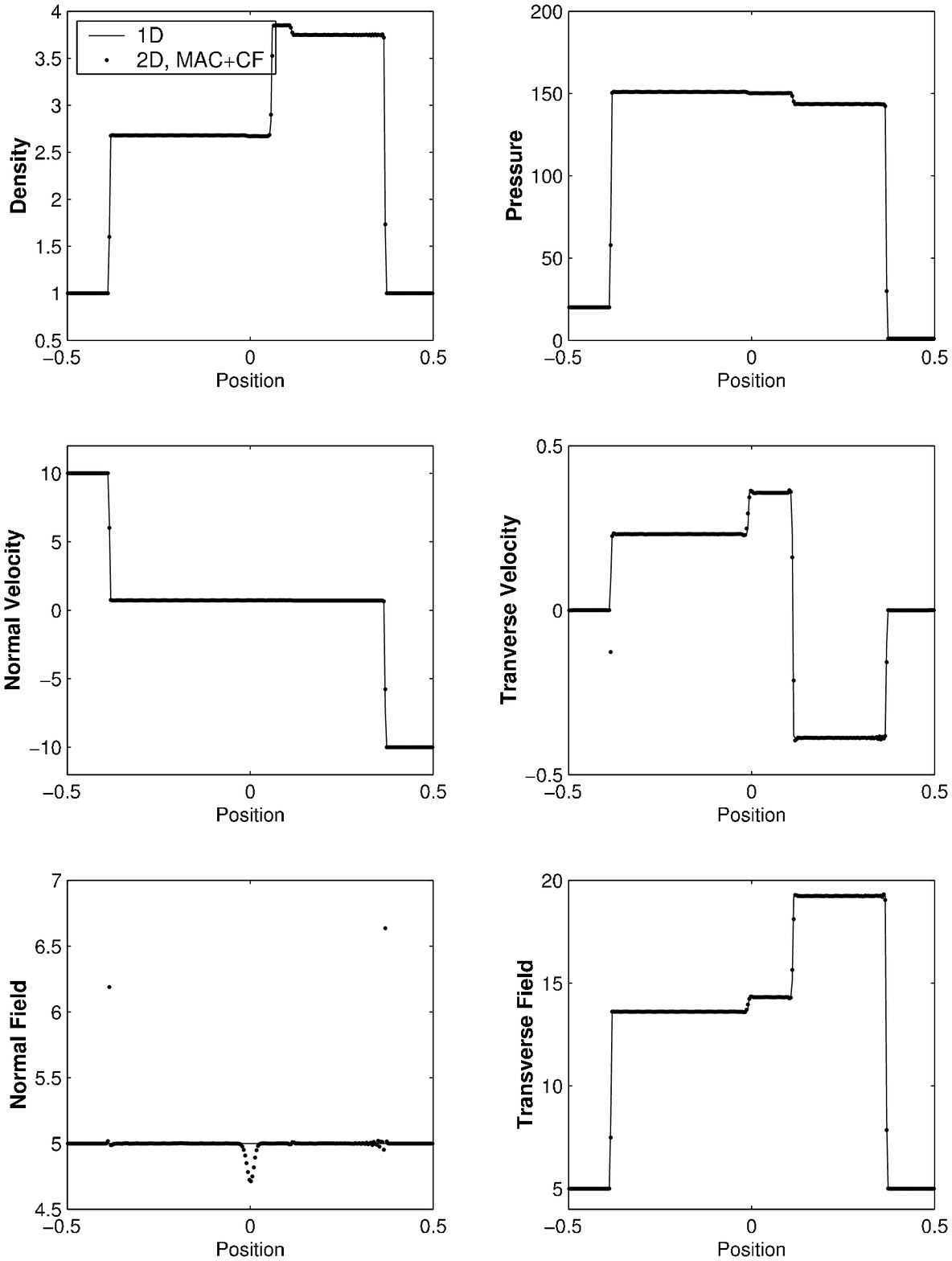}
\caption{MHD shock tube comparison by component. The 1-D shock tube
(line) is used as a basis for comparison with 2-D shock tube runs with
same effective resolution. The jumps in normal field $B_n$ are
expected, and the size of the jumps is similar to that published in
T\'oth 2000.}
\label{ist7.1}
\end{center}
\end{figure}
\par
\begin{figure}[!ht]
\begin{center}
\includegraphics{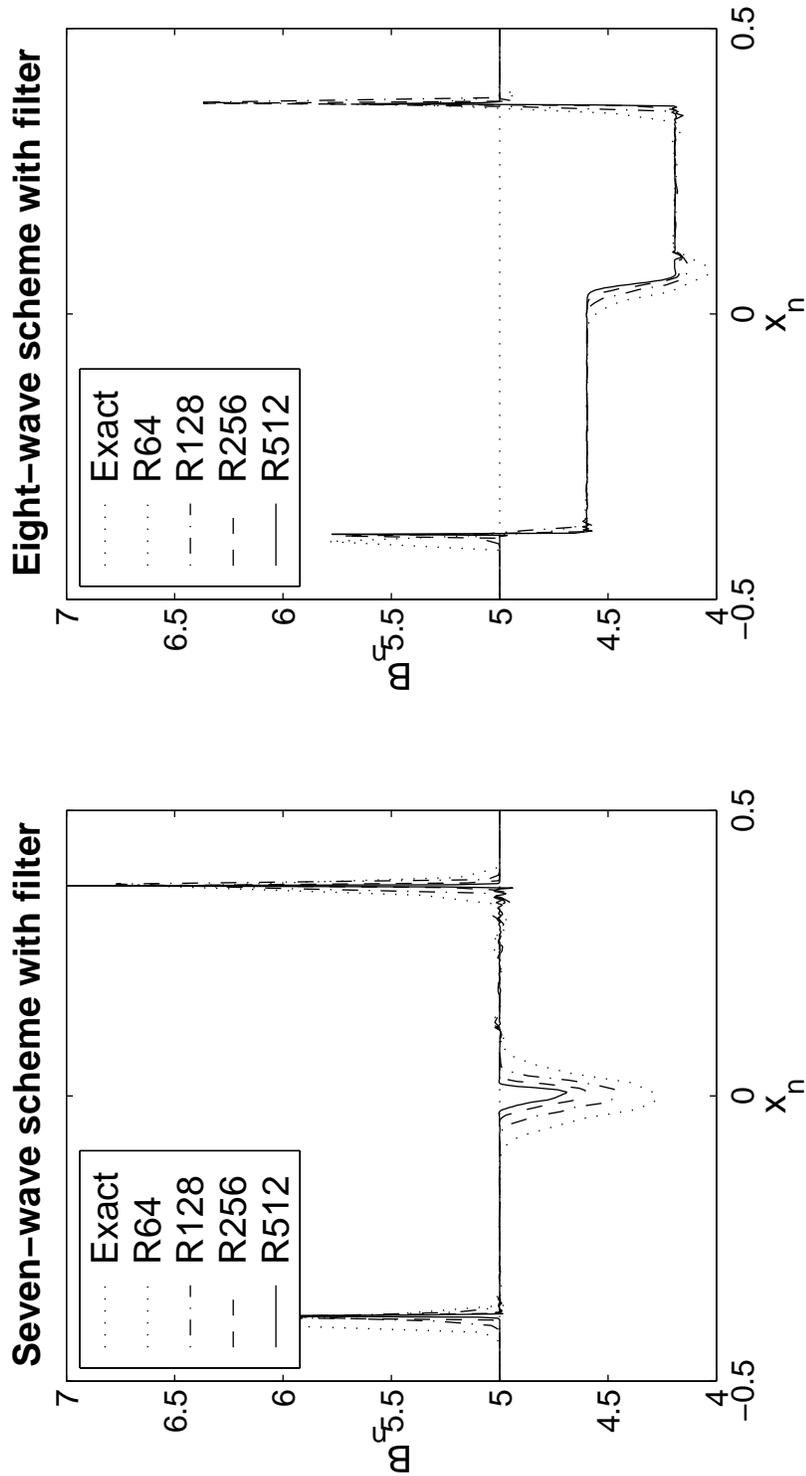}
\caption{Normal field results for MHD shock tube problem. Resolutions
from $R_{64}$ to $R_{512}$ are shown. Note the incorrect shock
structure in the eight-wave result.}
\label{istBnorm}
\end{center}
\end{figure}

Figure \ref{ist7.1} shows results for the inclined shock tube overlaid
on the (coarsened) aligned result. They compare a $R_{256}$ 1-D run
with a $N=256$ 2-D run that has equivalent effective grid spacing. The
ideal MHD codes shown give fractional errors with an 
$L^1$ norm of 3\% or less, averaged over all components. Errors in
the normal component of the field are relatively large, however. As
illustrated here and by Figure 11 in T00, shock-capturing codes
generally produce unphysical variation in the normal component of the
magnetic field inside the shock transition layer in non-grid-aligned
shock tube problems such as this. More generally, such codes exhibit
non-monotonic behavior inside the shock structure, so that Riemann
invariants are not exactly preserved there \cite{w86}. Importantly,
despite errors in $B_n$ inside the jump region, the jump relations are
still satisfied outside of it. This is clearly illustrated in Figure
\ref{istBnorm}, where we see the normal field for seven- and
eight-wave schemes with filtering. Note in particular that while the
eight-wave scheme converges to an answer, it does not give the correct
shock structure.
\par
Figures \ref{ist7.2} and \ref{ist8.2} show plots of the $L^1$ norm of
the error versus resolution for the 2-D shock tube runs. We first note
that neither the seven- nor eight-wave base code converges at first
order. This is most evident in the errors in the transverse velocity
and transverse field. The same plots indicate filtering alone is not
sufficient to produce reasonable convergence rates. A MAC projection
step added to either scheme produces much improved results (label
'MAC' in the figures), and the further addition of a filtering step
gives first-order convergence in all components (label
'MAC+CF'). The use of an approximate projection (labels 'CF+AP' and  
'MAC+CF+AP') produces decreased error in the normal field, indicating
a more accurate solution at given resolution. However, no improvement
in the convergence rates or errors in the other five components is
observed. When averaged over all components, the decrease in error
observed with an approximate projection is still noticeable, though
less so than indicated by considering the normal field alone.
\par
\begin{figure}[!ph]
\begin{center}
\includegraphics[height=0.8\textheight, width=0.92\textwidth,%
scale=1.0]{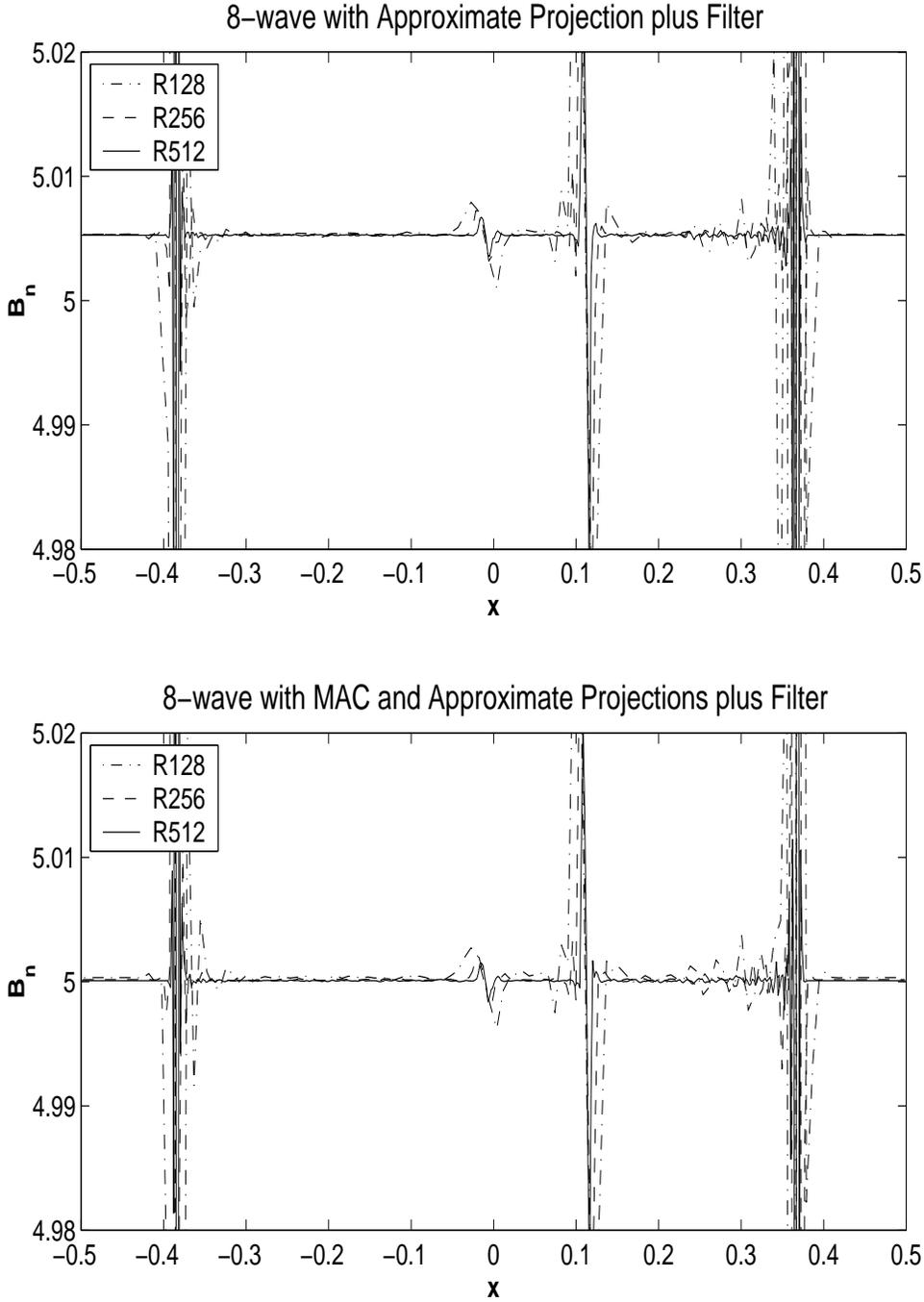}
\caption{Normal component of the field for eight-wave codes with
(bottom) and without (top) a MAC projection step. In these results for
the inclined shock tube, a cut along the shock propagation direction
is shown. The eight-wave code with the MAC projection converges to the
correct answer, $B_{normal}=5$. Without the MAC projection, the code
converges to an incorrect value for the normal field, about 0.1\%
higher than expected.}
\label{istBnormal}
\end{center}
\end{figure}
\begin{figure}[!ph]
\begin{center}
\includegraphics[height=0.92\textheight, width=0.92\textwidth,%
scale=1.0]{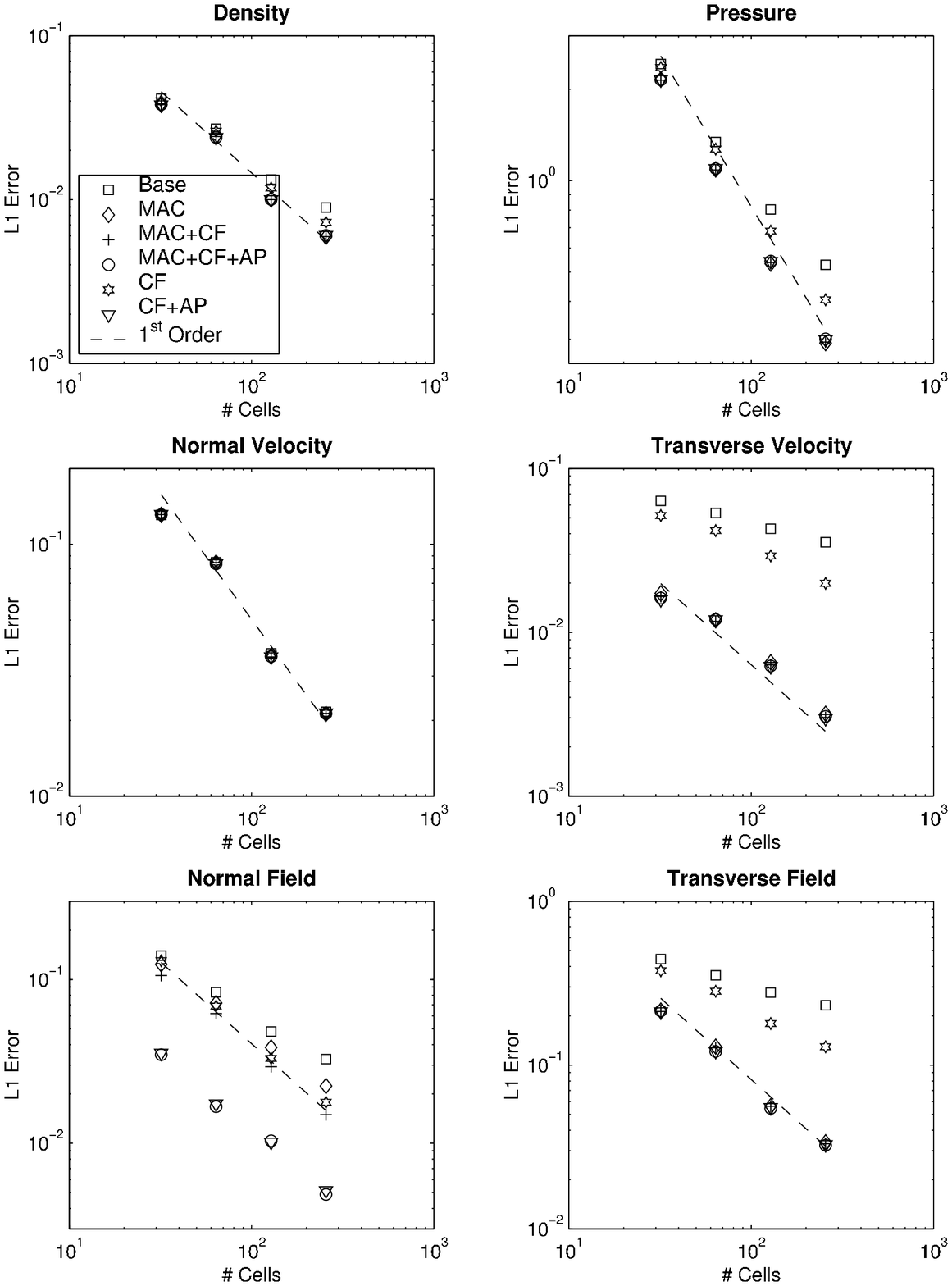}
\caption{$L^1$ norm of error versus resolution (number of cells per
dimension) for 2-D MHD shock tube run with variants of the seven-wave
code. The dashed line is a fiducial showing first-order fall-off of
errors. Note that some manner of projection step is required for
first-order convergence.}
\label{ist7.2}
\end{center}
\end{figure}
\begin{figure}[!ph]
\begin{center}
\includegraphics[height=0.92\textheight, width=0.92\textwidth,%
scale=1.0]{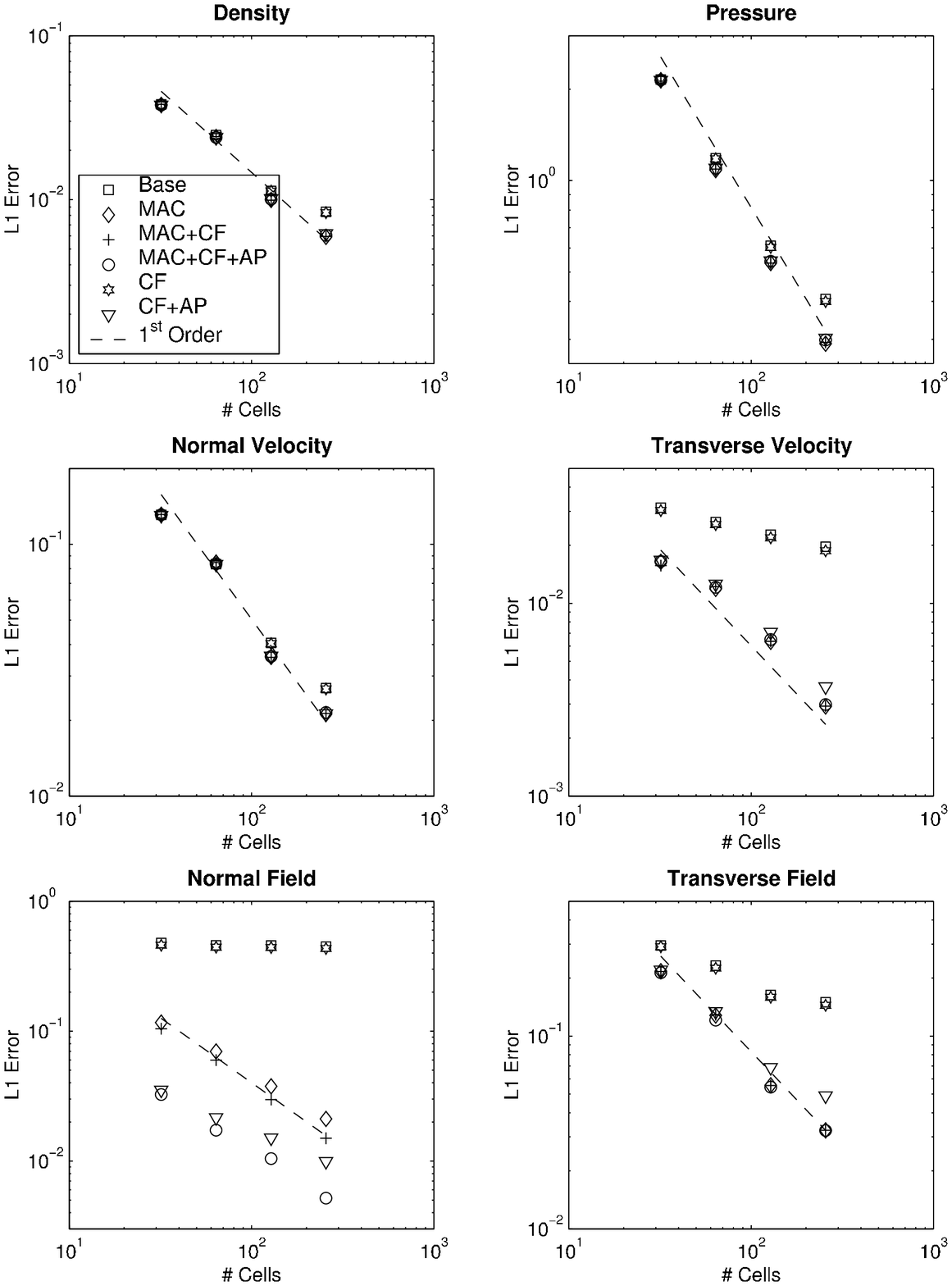}
\caption{$L^1$ norm of error versus resolution (number of cells per
dimension) for 2-D MHD shock tube run with eight-wave codes. The
dashed line is a fiducial showing first-order fall-off of errors.}
\label{ist8.2}
\end{center}
\end{figure}

While the results for seven- and eight-wave codes with a MAC
projection step are very similar, this picture changes somewhat when
the MAC projection is replaced with an approximate projection (label
'CF+AP'). The seven-wave code gives identical results whether or not a
MAC projection is used with the approximate projection and filter. The
eight-wave code still exhibits smaller errors when an approximate
projection is utilized. However, convergence of the errors for the
eight-wave code with approximate projection alone stalls at higher
resolution. The reason for this behavior becomes more clear upon
closer inspection of the normal field, plotted in Figure
\ref{istBnormal}. The normal field converges to an answer
that is 0.1\% above the expected value. This error, while masked by
the larger errors at the jumps at lower resolution, becomes
significant at higher resolution and consequently hampers
convergence. This error is observed to grow steadily over time.
\par

\subsection{Magnetized Flux Tube}
\label{aftResults}
This problem involves a high-field, low gas pressure region bounded on
both sides by a high-gas pressure, zero-field region. The physical
domain is of length $L_x=L_y=1$ on both sides. The base state has the
entire 2-D domain in pressure balance. The boundaries between
magnetized and unmagnetized regions are discontinuous and lie along
$x=\pm 0.2$. In both regions, $\rho=1$, $\vec{u}=0$ and
$B_{x}=B_{z}=0$ initially. In the magnetized region, $x\in
[-0.2,0.2]$, $B_{y}=\sqrt{80\pi}$ and $P=1$. Outside of this, $B_y=0$
and $P=11$. To the background state is added a sinusoidal perturbation
upon the x-velocity whose amplitude is $\delta_{\rm pert}=0.01$ times
the Alfv\'en speed, and covers the entire domain: 
\begin{equation}
\delta u = \delta_{\rm pert}c_A\sin\left(2\pi y\right) \;\; %
\textrm{and}\;\; \delta \vec{B} = \delta\rho = \delta P = \delta v = %
\delta w = 0
\end{equation}
\par
The Alfv\'en speed in this case is
$c_A=B_0/\sqrt{4\pi\rho}=\sqrt{20}$. The strong discontinuity in the
field is expected to cause problems for algorithms that do not
suppress non-solenoidal fields. In particular, at the stagnation
points where the perturbation velocity is zero, truncation errors
leading to non-solenoidal fields can build up and cause numerical
schemes to go unstable. This is in fact what we find for base schemes
run without a projection step. Non-solenoidal fields build up at
these stagnation points, causing the production of spurious velocities
and triggering numerical instabilities.
\begin{figure}[!ph]
\begin{center}
\includegraphics[height=0.88\textheight, width=0.92\textwidth,%
scale=1.0]{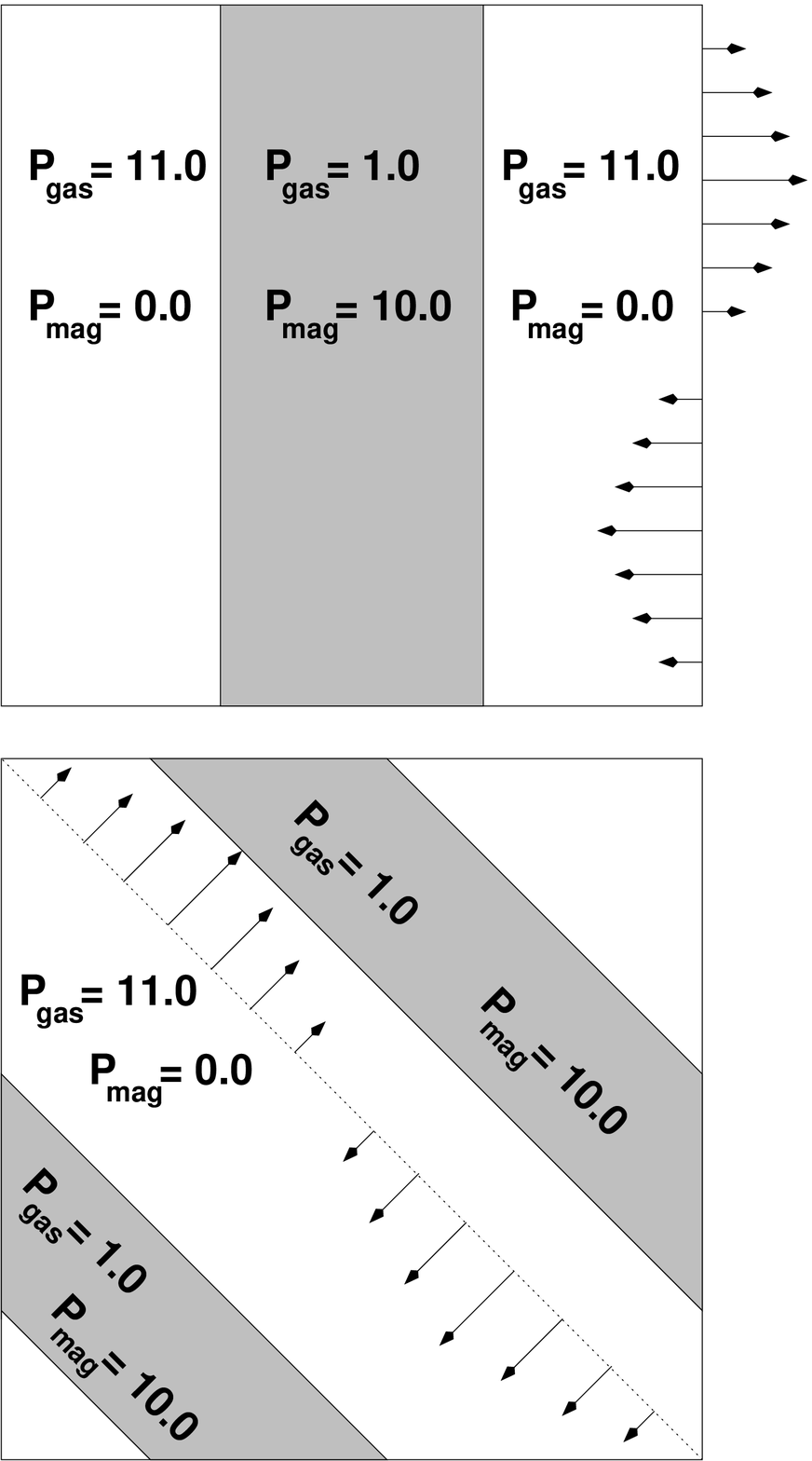}
\caption{Grid-aligned (top) and inclined (bottom) flux tube initial
conditions. Vectors indicate the perturbation velocity field; it is
constant along lines parallel to the vectors. There are two regions:
one magnetized, the other not. The entire domain is initially in total
(magnetic plus thermal) pressure balance. The velocity perturbation at
1\% of the Alfv\'en speed is applied to the entire domain, causing the
tube and surrounding unmagnetized medium to oscillate.} 
\label{aftIC}
\end{center}
\end{figure}
\par
Figure \ref{aftIC} shows the initial conditions for the flux tube
problem. The perturbation plucks the field. After an initial transient
at start-up, the perturbation develops into a standing wave in the
magnetized region. The unmagnetized region sloshes back and forth from
one side of the flux tube to the other, owing to the periodic boundary
conditions. We expect the initial velocity perturbation to cause
standing Alfv\'enic waves in the magnetized region and, because of the
slightly different oscillation frequencies of the two regions, compressive 
waves at the boundaries of the tube. The change in the internal energy
tracks the compressive waves. The problem was run to a time $t=6.0$,
corresponding to about 60 Alfv\'en crossings of the short dimension of
the tube. This was enough time for the problem to exhibit stable
periodic behavior and subsequently evolve for many periods of this
oscillation.
\begin{figure}[!ph]
\begin{center}
\includegraphics[height=0.92\textheight, width=0.92\textwidth,%
scale=1.0]{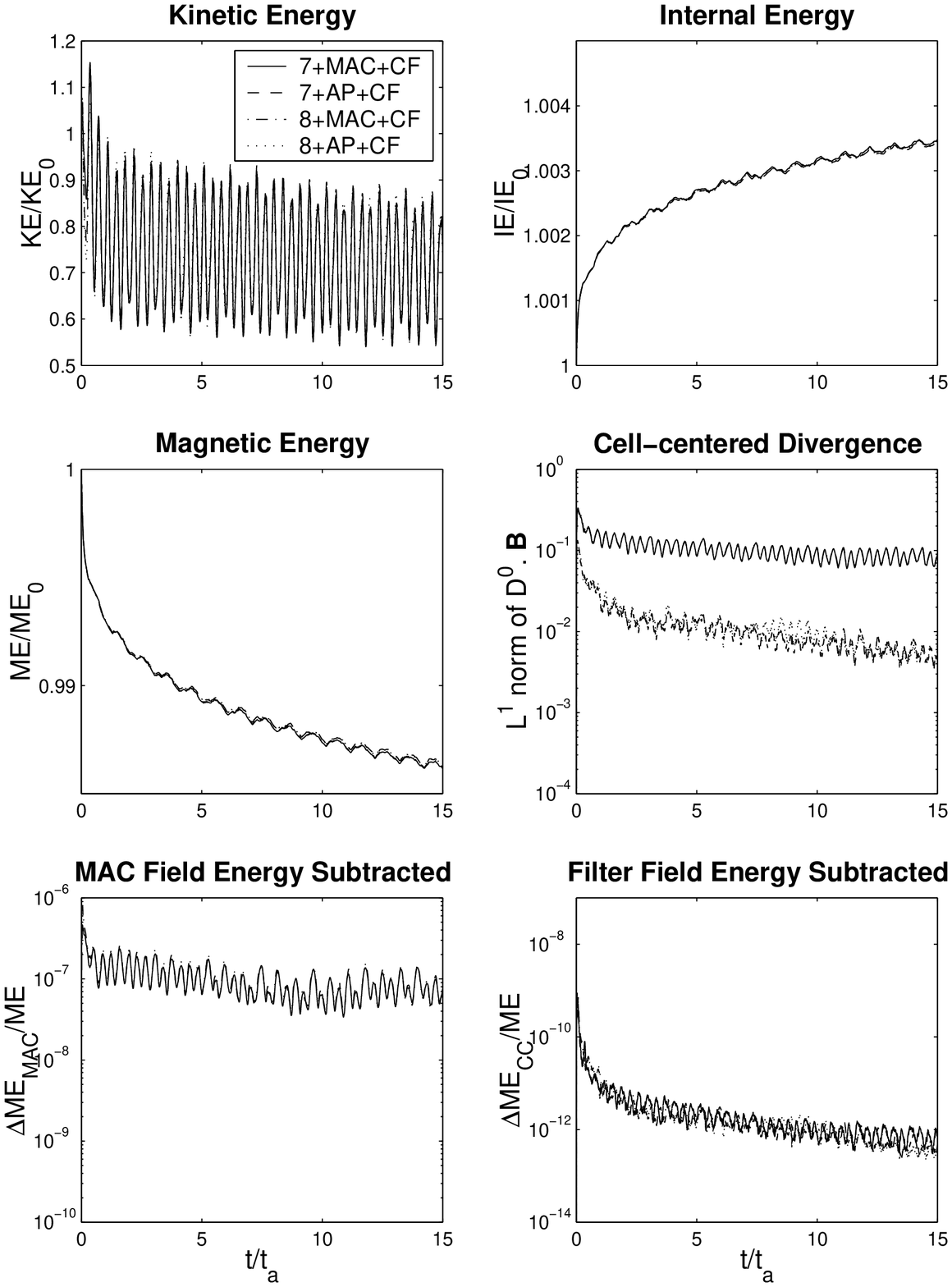}
\caption{Grid-aligned flux tube results at $R_{512}$. We use the $L^1$
norm of $\nabla\cdot\vec{B}$, plus total energies, to follow the 
evolution. Energies are normalized to their initial values. All
quantities are plotted in units of the Alfv\'en crossing time,
$t_a=1/c_a=0.2236$. Despite their different means of suppressing
monopoles, the codes performed very similarly on this test.}
\label{aft}
\end{center}
\end{figure}
\par
In this case we tested both seven- and eight-wave codes, but only with
either a MAC projection and filter, or an approximate projection and
filter. This is justified by the sub-par performance of the other
variants on the inclined shock tube problem. These four versions of
the code performed almost identically on this test, as evidenced by
Figure \ref{aft}. There, we plot global quantities such as kinetic and 
magnetic energy and $L^1$ norm of the monopole density. Note that the
cell-centered divergence of the field is smaller for the approximate
projection, though the dynamics remain the same.
\par

\subsection{Inclined Flux Tube}
A second version of the flux tube problem, in which it is inclined at
an angle of 45 degrees with respect to the original, is better at
differentiating between the algorithms. It constitutes a strong test
of the robustness and stability of the codes.
\par
The size of the domain was $L_x=L_y=\sqrt{2}$, and both magnetized and
unmagnetized regions have the same physical extent as in the aligned
case. In both regions, we have $\rho=1$, $\vec{u}=0$ and $B_{z}=0$
initially. In the magnetized region, $B_x=-\sqrt{40\pi}$,
$B_y=\sqrt{40\pi}$, and $P=1$. In the unmagnetized region, $P=11$ and
$B_x=B_y=0$. The perturbation is again applied to the entire domain,
and has strength $\delta_{\rm pert}=0.01$:
\begin{eqnarray}
\delta u = \delta_{\rm %
pert}\frac{c_A}{\sqrt{2}}\sin\left(2\pi\frac{-x+y}{\sqrt{2}}\right) \\ 
\delta v = \delta_{\rm %
pert}\frac{c_A}{\sqrt{2}}\sin\left(2\pi\frac{-x+y}{\sqrt{2}}\right) \\ 
\delta \vec{B} = \delta\rho = \delta P = \delta w = 0
\end{eqnarray}
One effect of rotating the flux tube is the addition of a numerical
perturbation, or gridding effect. The volume-averaging required for
producing the initial conditions creates a region of intermediate
pressure and field between the fully magnetized and unmagnetized
regions. The total pressure in this intermediate region was kept the
same as elsewhere in the domain. The physical extent, and presumably
physical effects, of this transition region become smaller as the
resolution is increased. The second effect of our rotation is that the
physical domain was somewhat larger. As mentioned above, this was done
in order to keep the physical width of the magnetized and unmagnetized
regions the same as for the aligned flux tube. The result is a factor
of $\sqrt{2}$ mismatch between the grid spacing in the aligned and
inclined flux tube runs.
\par
We expect many qualitative similarities between the aligned and
inclined flux tube results. The characteristics of the start-up
transients are somewhat different, due in part to the gridding
effects. The slightly different grid spacing also means that slightly
higher resolution is required to obtain comparable results. Indeed,
these tests were run at resolutions ranging by factors of two from
$R_{128}$ to $R_{1024}$, a factor of two higher than for the aligned
case.
\begin{figure}[!ph]
\begin{center}
\includegraphics[height=0.82\textheight, width=0.92\textwidth,%
scale=1.0]{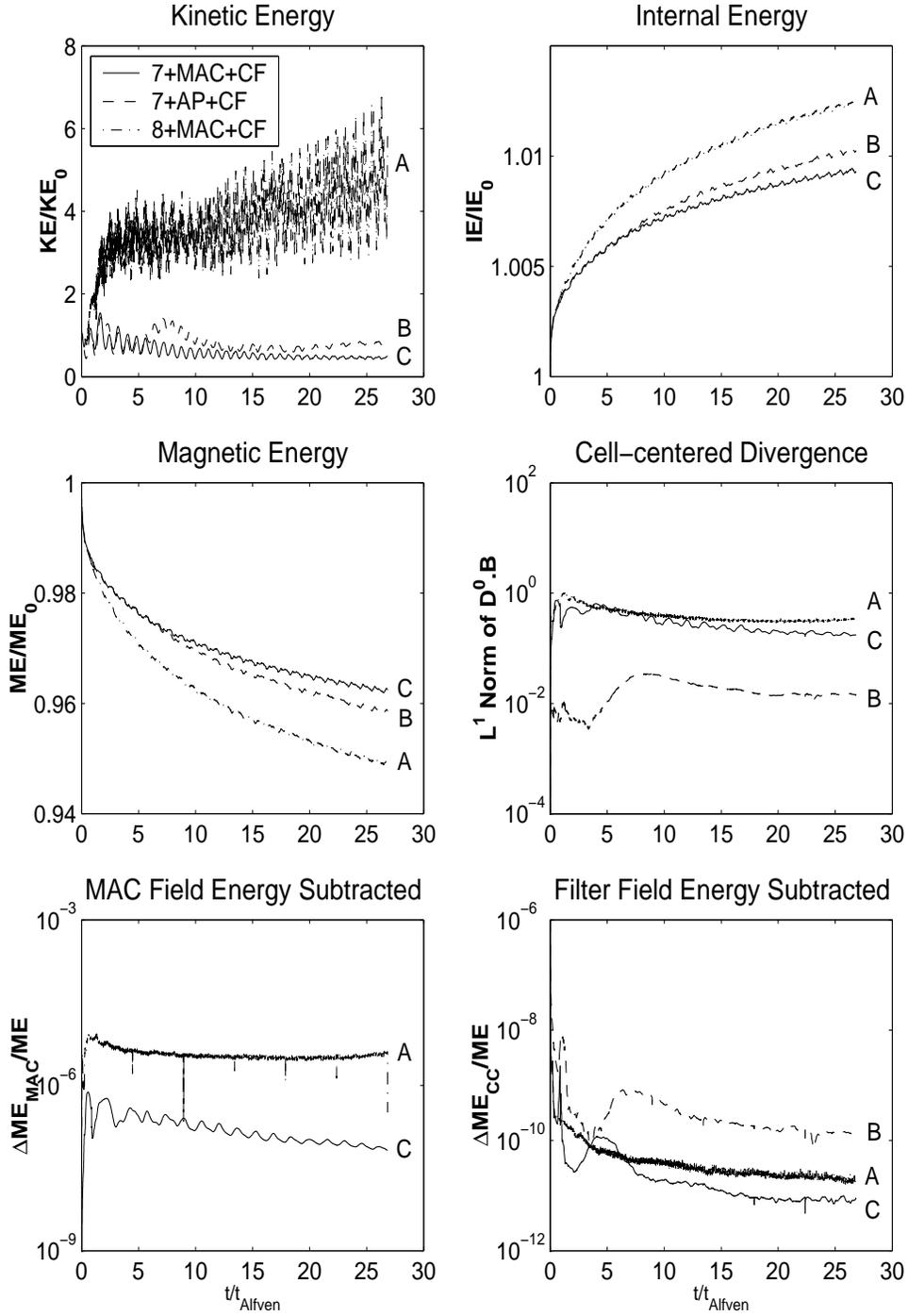}
\caption{Inclined flux tube results for one eight-wave (``8+MAC+CF'')
and two seven-wave (``7+AP+CF'' and ``7+MAC+CF'') codes. The results
for the different codes are also labeled with 'A', 'B', and 'C',
respectively. Since a MAC projection is not used in 7+AP+CF ('B'), no
field energy is subtracted and it does not appear in the graph at
lower-left. Note the relative size of increases in the kinetic energy
of the codes, with smaller kinetic energies indicating stable
oscillations. The eight-wave code with approximate projection and
filter (``8+AP+CF'') was run on this problem as well, but became
unstable.}
\label{ift7}
\end{center}
\end{figure}
\par
We again chose to run only those codes that showed close to
first-order convergence for the inclined shock tube on this
problem: seven- and eight-wave codes with either MAC projection and
filter, or approximate projection and conservative filter. There was
very little difference between the two seven-wave codes. The
difference between seven- and eight-wave codes was more
noticeable. The eight-wave codes both experienced large increases 
in kinetic energy, indicating that the magnetized tube was no longer
in equilibrium. The size of this increase was largest ($\sim
10^3\rm{KE}_{\rm initial}$) for the eight-wave code with approximate
projection, which also experienced substantial deviations from energy
conservation. The size of these kinetic energy jumps decreased with
resolution.
\begin{figure}[!ph]
\begin{center}
\includegraphics[height=0.92\textheight, width=0.92\textwidth,%
scale=1.0]{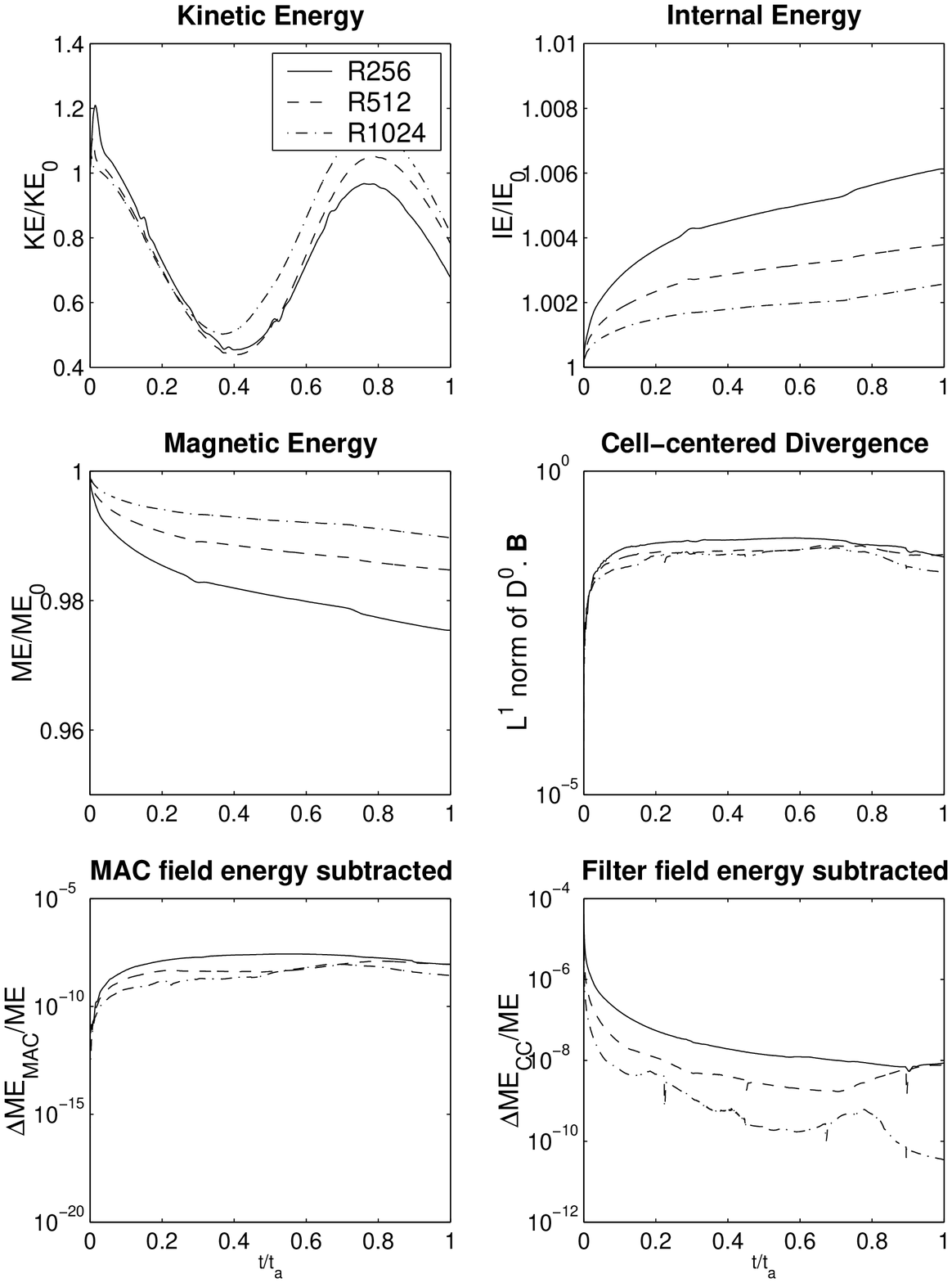}
\caption{Inclined flux tube results for 256, 512, and 1024 cells in
each direction. These results used the 7-wave MHD code with
filtering and a coefficient of 0.1. The quantities
shown are the same as in Figure \ref{aft}.}
\label{ift7res}
\end{center}
\end{figure}
\par
Figure \ref{ift7res} shows a resolution study of the inclined flux
tube, with global quantities plotted out to $t=1$. (This is about 5
Alfv\'en crossing times; a shorter time was chosen in order to show
more detail. All codes that were stable to this time were also stable
out to $t=6$.) The plots include kinetic, internal, and magnetic
energies, all scaled to their initial values. We also show the
$L^1$ norm of cell-centered measure of the divergence,
$D^0\cdot\vec{B}$, along with the magnetic energy change due to
the MAC projection and filtering steps. First, note that the internal
and magnetic energies appear to be converging, while convergence of
the kinetic energy is not so clear. This result can be attributed to
the large difference in the size of these quantities; the initial
kinetic energy is $\sim 10^5$ times smaller than the initial internal
energy, and $\sim 10^4$ times smaller than the initial magnetic
energy. Thus, small errors in either of these quantities can appear as
sizeable changes in the kinetic energy. We note in particular that
increases in the kinetic energy seem to correlate with increases in
the fraction of the magnetic energy subtracted in either the filter or
the MAC projection.
\par
The expected qualitative similarities between aligned and inclined flux
tube results are present, such as oscillations in kinetic and magnetic
energy of similar magnitudes. The size of the oscillations in the
$L^1$ norm of $v_t$ and $v_n$ were also found to be very
similar. However, substantial differences are also present. Comparing
Figures \ref{ift7} and \ref{aft}, we note that the transient rise in
kinetic energy early in the simulation begins later for the inclined
tube than for the 1-D tube. In fact, the rise begins later and has
smaller amplitude as resolution is increased, a fact that we attribute
in part to the decrease in physical extent of the transition region. 
\begin{figure}[!ht]
\begin{center}
\includegraphics[height=14cm, width=14cm,%
trim=0cm 0cm 0cm 0cm,clip]{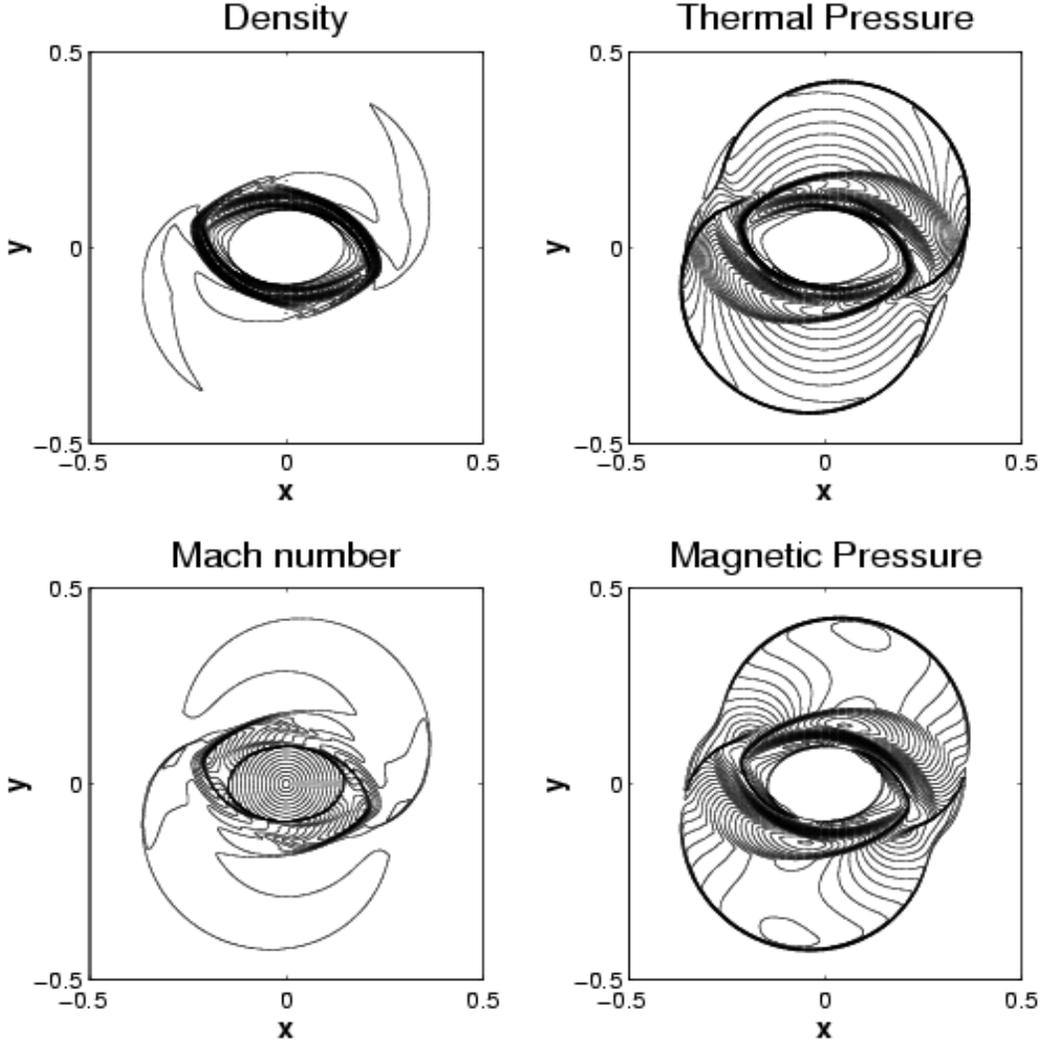}
\caption{Rotor problem results reproducing Figure 18 from T00. The
plot has the same number of contours and these contours lie between
the same limits. A slightly expanded domain was used ($x,y\in
[-0.64,0.64]$) in order to preserve the same grid spacing.}
\label{rotor}
\end{center}
\end{figure}
\par
As a final test of the nonlinear behavior of the code, we reproduce the
first MHD rotor problem outlined in T00, earlier performed by Balsara
and Spicer \cite{bs99}. This problem constitutes a central,
high-density ($\rho=10.0$) region surrounded by a low-density
medium. The central region is given a constant angular velocity. The
entire domain is threaded by an initially constant field,
$\vec{B}=5.0\hat{e}_x$. The rotation winds the magnetic field, sending
Alfv\'en waves propagating into the surrounding medium. 
\par
This test was run using both seven-wave codes (MAC+CF and AP+CF) and
one eight-wave variant (MAC+CF). We see no negative pressures in any
of the code variants, and obtain first-order convergence (errors
decreasing as the grid spacing to the 0.95 power, as averaged over
all variables) using the seven-wave codes, and the eight-wave code
with a MAC projection. A reproduction of Figure 18 from T00 is given
in Figure \ref{rotor}. Note the steeper gradient in Mach number in
comparison to the T00 result; we do not see the peaks therein attributed
to pressure undershoots.

\section{Conclusions}
We have presented an unsplit method for ideal MHD which, when combined
with projection and filtering steps, shows no effects of
non-solenoidal fields, while retaining the co-location of all physical
quantites at cell-centers. The latter point is important because such
a uniform centering makes it easier to extend the scheme to adaptive
meshes, and has the advantage of using a well-understood Godunov
method for time integration. These are in contrast to the Constrained
Transport (CT) approach, where staggered grids add additional software
complexity. Furthermore, it is not obvious how to discretize diffusion
operators to include non-ideal MHD effects on staggered grids. Lastly,
we add that the effective advection scheme for the magnetic field
components is not of the standard type, and it is unclear what its
properties are in the presence of under-resolved gradients.
\par
The complexity and cost of the scheme relative to others is an
important consideration. Our scheme requires three characteristic
analysis steps plus twelve Riemann solution steps in 3-D. This is
comparable to the six characteristic analysis and six Riemann
solution steps required in a two-step Runge-Kutta scheme. The
computational cost of projection is not insignificant, and adds
additional software complexity. On a single grid and a single
processor, the cost of the projection when computed using FFTs is a
small fraction of the overall cost of the computation (less than 200
floating-point operations per cell in 3D). Current research on
analysis-based solvers for Poisson's equation \cite{hg00,bc02} have
the potential to make the cost of this remain less than that of the
hyperbolic calculation, on both adaptive grids and on parallel
processors.
\par
The Hodge projection of fields either at cell-edges (MAC projection)
or cell-centers (in our case via an approximate projection) was found
to be essential to accuracy and stability. The MAC projection was
essential for accuracy of the eight-wave code in the presence of
discontinuities. The seven-wave code results on the inclined shock
tube were less sensitive to the type of projection used. The inclined
flux tube results painted essentially the same picture, though the
seven-wave MAC projected result was somewhat more robust.
\par
Use of a filter alone on linear problems reduces computational cost
and does not affect accuracy. We find that both filter and projection
are required on strongly nonlinear problems such as the 
inclined flux tube. Also essential to the accuracy of the scheme are
the modifications suggested by Stone -- linear waves not propagating
along coordinate-axis directions are not second-order accurate without
it.
\par
We found that in determining the accuracy of our code, both measuring
the rate of convergence on nonlinear problems and comparing absolute
magnitude of the error were required. Significant differences
were found between the convergence rates even of individual primitive
variables. These data can be crucial in choosing between different
algorithmic variants of a base code. We encourage future authors to
incorporate convergence testing, taking care to calculate errors in
each variable separately, of simple nonlinear problems such as the
shock tubes run here into their test suites. 
\par
Overall the ideal MHD code with a projection, either MAC or
approximate, and a filter performed best on the suite of
tests presented. Filtering of the magnetic field was important to code
stability in the nonlinear problems, and accuracy in the deficient
wave problems.
\par
The magnetized, perturbed flux tube constitutes a strong test of the
stability of our schemes. The combination of a stationary
discontinuity, low ratio of thermal to magnetic pressure (beta), and
imposed perturbation caused significant problems for the base unsplit
code without projection or filter. With a suitably chosen value of the
filtering coefficient, the filter helped stability and
worked to decrease the magnitude of a centered-difference measure of
the divergence of the field.
\section{Acknowledgements}
The authors would like to thank Jim Stone for pointing out the missing
multidimensional MHD terms.
\par
The research of Robert K. Crockett was supported in part by A Division
at Lawrence Livermore National Laboratory.
\par
The research of Christopher F. McKee was supported in part by NSF
grant AST-0098365.
\par
Work at the Lawrence Berkeley National Laboratory is sponsored by the
US Department of Energy Applied Mathematical Sciences program under
contract DE-AC03-76SF00098 and by the NASA Earth and Space Sciences
Computational Technologies Program under interagency agreement number
S-44830-X.
\par
The research of Richard I. Klein was supported in part by a NASA ATP
grant NAG5-12042. Both Richard I. Klein and Robert T. Fisher are
supported under the auspices of the US Department of Energy at the
Lawrence Livermore National Laboratory under contract W-7405-ENG-48.

\appendix
\section{Appendix A: Divergence Constraints, Modified Equation
Analysis, and Eigenvector Deficiencies}
\label{def}
In this section, we will present a heuristic analysis of the effect of
numerical errors in the divergence-free constraint on the stability of
finite-difference methods for the ideal MHD equations. Our starting
point will be the modified equation approach to analyzing the effect
of truncation error on solution error. For any finite difference
method, the modified equation is given by the original system of PDEs,
with forcing terms given by the truncation error. In the present
setting, the modified equation takes the following form:
\begin{eqnarray}
\partial_t U^{Mod} + \nabla\cdot \vec{F}(U^{Mod}) = \tau_U(U^{Mod}) \\
\nabla\cdot \vec{B}^{Mod} = \tau_D(U^{Mod}).
\end{eqnarray}
Here $\tau_U$ is the usual truncation error for the numerical method
obtained from applying the difference operator to a solution to the
differential equation evaluated on the grid. The equation for the
evolution of $\tau_D$ is obtained by taking the divergence of the
modified equation for the evolution of $U^{Mod}$,
\begin{equation}
\partial_t  \tau_D = \nabla\cdot \tau_B.
\end{equation}
The truncation error forcing terms mimic the effect of numerical error
on the computed solution. Specifically, we expect $U^{Mod}$, the
solution to the modified equation, to satisfy $||U^{\Delta
x}-U^{Mod}|| = \mathcal{O}(\Delta x^{p+1})$, where $U^{\Delta x}$, the
solution obtained from the $p^{\rm th}$-order scheme on a grid with
spacing $\Delta x$, satisfies $||U^{\Delta x}-U||=\mathcal{O}(\Delta
x^p)$.
\par
In particular, for MHD, the effect of numerical error can be
understood in terms of the truncation-error forcing in the modified
equation causing the solution to violate the divergence-free
constraint. Without that constraint being satisfied, the remaining
ideal MHD equations can exhibit eigenvector deficiencies in the
linearized-coefficient matrix $\mathbf{A}$, leading to anomalous loss
of regularity and ill-posedness. In the numerical simulation, this
translates into loss of accuracy and possibly instability of the
underlying difference method.
\par
To see this, we consider the case of a small-amplitude wave
corresponding to one of the eigenmodes of $\mathbf{A}$ (see Equation
\ref{charInterp}):
\begin{eqnarray}
W(\vec{x},\, t) = W_0 + \alpha(x-\lambda_k t)r_k \\
\mathbf{A}_0r_k = \lambda_k r_k \;\; r_k = (\tilde{r}_k,\, 0)^T.
\end{eqnarray}
Then $W(\vec{x},\,t)$ satisfies the MHD equations up to terms of
$\mathcal{O}(\alpha^2)$.
\par
Without loss of generality, we take the direction of propagation to be
in the x-direction in 3-D. However, we allow our computational spatial
grid to have an arbitrary orientation in space. In that case, the
modified equation corresponding to our numerical solution to the PDE
in primitive variables $W$ is given by (see Equation \ref{lagrSplit})
\begin{eqnarray}
\partial_t W^{Mod} + \mathbf{A}_0 \partial_x W^{Mod} = \tau_W \\
W^{Mod} = (\tilde{W}^{Mod},\, B_x^{Mod}).
\end{eqnarray}
If we define the new variable $\alpha^{Mod} = l_k\cdot
(\tilde{W}^{Mod} - \tilde{W}_0)$, the modified equation dynamics can
be reduced to the following system of two equations:
\begin{eqnarray}
\label{modEvolAlpha}
\partial_t \alpha^{Mod} + \lambda_k \partial_x\alpha^{Mod} + (l_k\cdot %
a_B)\partial_xB_x^{Mod} = l_k\cdot\tilde{\tau} \\
\label{modEvolE}
\partial_t B_x^{Mod} = \tau_B 
\end{eqnarray}
In the system here, if one of the computational spatial grid
coordinate axes is aligned with $\hat{x}$, $\tau_B \equiv 0$. However,
if the direction of propagation is not aligned with one of the
computational coordinates, then in general $\tau_B\neq 0$. If
$l_k\cdot a_B \neq 0$ and $\lambda =0$, then the left-hand side of
\ref{modEvolAlpha}-\ref{modEvolE} is an example of a first-order system
with an eigenvector deficiency. Such systems have an obvious loss of
spatial regularity: $\alpha^{Mod}$ grows like the derivative of
$\tau_B$. This is in contrast to the behavior of well-posed hyperbolic
systems, in which the solution has the same spatial regularity as the
forcing. In other words, since regularity implies that there are as
many derivatives in the solution as in the forcing term, discontinuous
forcing of a hyperbolic system leaves the problem
ill-posed. Discontinuity in $\tau_B$ implies that $\alpha^{Mod}$ can
grow without bound, owing to its dependence on the derivative of
$\tau_B$.
\par
In terms of a numerical method, we expect that the presence of such
terms would lead to either an anomalous loss of accuracy or numerical
instability. In the latter case, the forcing of $\alpha^{Mod}$ in
\ref{modEvolAlpha} takes the form of a finite difference operator
applied to $B_x^{Mod}$, whose spatial variation is due entirely to
$\tau_B$. If $\tau_B$ fails to be smooth, either because of lack of
smoothness in the initial data or in the finite difference formulae
(eg. limiters), such lack of smoothness is immediately amplified.
\par
This discussion also provides an explanation for the behavior of the
method described here. The use of the MAC projection and the filter
does not eliminate the truncation error terms that lead to the
eigenvector deficiency, but regularizes it by smoothing. For example,
the application of the filter in the plane-wave example corresponded
to adding a diffusion term to the equation for $B_x^{Mod}$,
\begin{equation}
\partial_t B_x^{Mod} = \tau_B + \eta \partial_x^2 B_x^{Mod},
\end{equation}
with $\eta = \mathcal{O}(\Delta x)$. The use of the filter alone is
sufficient to stabilize the small-amplitude plane wave solution in
Section \ref{waves}, and in that case leads to a second-order accurate
result. The MAC projection performs a more drastic smoothing, but only
on the intermediate form of $B_x^{Mod}$ used to compute $\partial_x
B_x^{Mod}$ in equation \ref{modEvolAlpha}.

\end{document}